\documentclass[10pt,doublecolumn,twoside]{IEEEtran}

\usepackage{etoolbox}
\newtoggle{doublecolumn}
\newtoggle{calculateFigures}

\toggletrue{doublecolumn}
\togglefalse{calculateFigures}

\newcommand\B{\boldsymbol}
\newcommand\C{\mathcal}
\newcommand\M{\mathbb}

\usepackage{etex}

\usepackage[english]{babel}
\usepackage{lipsum}
\usepackage{amsbsy,amsmath,amsfonts,amssymb,amsthm}
\usepackage{mathtools}
\usepackage{textcomp} 
\usepackage{relsize}

\usepackage{bm,cite,cases,pstricks,times,url,verbatim} 
\usepackage[noend]{algpseudocode}
\usepackage{booktabs}
\usepackage{tabularx,array,dcolumn,multirow}

\usepackage{yfonts}

\usepackage{dutchcal} 

\usepackage{soul} 

\usepackage{blkarray,bigdelim} 

\allowdisplaybreaks[4]

\usepackage{centernot} 

\newcommand{\subparagraph}{}

\iftoggle{doublecolumn}{
    \newtheorem{thm}{Theorem}
    \newtheorem{fact}{Fact}
    \newtheorem{lemma}{Lemma}
    \newtheorem{definition}{Definition}
    \newtheorem{conj}{Conjecture}
    \newtheorem{propos}{Proposition}
    \newtheorem{corol}{Corollary}
    \newtheorem{ass}{Assumption}
    \newtheorem{example}{Example}
    \newtheorem{remark}{Remark}
    \newtheorem{note}{Note}
    \newtheorem{obs}{Observation}
}{

    \newtheoremstyle{exampstyle}
      {0} 
      {0} 
      {\itshape} 
      {} 
      {\bfseries} 
      {.} 
      {.5em} 
      {} 

    \theoremstyle{exampstyle} \newtheorem{thm}{Theorem}
    \theoremstyle{exampstyle} 
    \theoremstyle{exampstyle} \newtheorem{lemma}{Lemma}
    \theoremstyle{exampstyle} \newtheorem{definition}{Definition}
    \theoremstyle{exampstyle} 
    \theoremstyle{exampstyle} \newtheorem{propos}{Proposition}
    \theoremstyle{exampstyle} \newtheorem{corol}{Corollary}
    \theoremstyle{exampstyle} 
    \theoremstyle{exampstyle} \newtheorem{example}{Example}
    \theoremstyle{exampstyle} \newtheorem{remark}{Remark}
    \theoremstyle{exampstyle} 
    \theoremstyle{exampstyle} 
}

\newcommand{\argmax}[1]{\underset{#1}{\operatorname{arg}\,\operatorname{max}}\;}

\makeatletter
\newcommand{\pushright}[1]{\ifmeasuring@#1\else\omit\hfill$\displaystyle#1$\fi\ignorespaces}
\newcommand{\pushleft}[1]{\ifmeasuring@#1\else\omit$\displaystyle#1$\hfill\fi\ignorespaces}

\begingroup
\catcode`\#=11
\gdef\noautorotate{-dAutoRotatePages#/None}
\endgroup

\usepackage{graphicx,xcolor,float,dblfloatfix}
\usepackage{psfrag}

\usepackage{caption}
\usepackage{subcaption}

\newcommand{\subalign}[1]{%
  \vcenter{%
    \Let@ \restore@math@cr \default@tag
    \baselineskip\fontdimen10 \scriptfont\tw@
    \advance\baselineskip\fontdimen12 \scriptfont\tw@
    \lineskip\thr@@\fontdimen8 \scriptfont\thr@@
    \lineskiplimit\lineskip
    \ialign{\hfil$\m@th\textstyle##$&$\m@th\textstyle{}##$\crcr
      #1\crcr
    }%
  }
}

\iftoggle{calculateFigures}{
    \usepackage[crop=off,runs=2,pspdf=\noautorotate]{auto-pst-pdf}
}{
    \usepackage[off]{auto-pst-pdf}
}

\usepackage{hyperref}

\graphicspath{%
{./Figures/}
}

\usepackage{caption}
\begin{document}

\author{\IEEEauthorblockN{Alessandro~Biason,~\IEEEmembership{Student~Member,~IEEE,} Nicola~Laurenti, and~Michele~Zorzi,~\IEEEmembership{Fellow,~IEEE}
\thanks{The authors are with the Dept. of Information Engineering, University of
Padova, Padova, Italy. email: \{biasonal,nil,zorzi\}@dei.unipd.it.}
\thanks{A preliminary version of this paper has been presented at IEEE GLOBECOM 2015~\cite{Biason2015b}.}
}}

\title{Achievable Secrecy Rates \\ of an Energy Harvesting Device}

\maketitle
\thispagestyle{empty}
\pagestyle{empty}

\begin{abstract}
The secrecy rate represents the amount of information per unit time that can be securely sent on a communication link. In this work, we investigate the achievable secrecy rates in an energy harvesting communication system composed of a transmitter, a receiver and a malicious eavesdropper. 
In particular, because of the energy constraints and the channel conditions, it is important to understand when a device should transmit and to optimize how much power should be used in order to improve security. Both full knowledge and partial knowledge of the channel are considered under a Nakagami fading scenario. We show that high secrecy rates can be obtained only with power \emph{and} coding rate adaptation. Moreover, we highlight the importance of optimally dividing the transmission power in the frequency domain, and note that the optimal scheme provides high gains in secrecy rate over the uniform power splitting case. Analytically, we explain how to find the optimal policy and prove some of its properties. In our numerical evaluation, we discuss how the maximum achievable secrecy rate changes according to the various system parameters. Furthermore, we discuss the effects of a finite battery on the system performance and note that, in order to achieve high secrecy rates, it is not necessary to use very large batteries.
\end{abstract}


\section{Introduction}

\IEEEPARstart{S}{ecurity} and privacy are becoming more and more important in communications and networking systems, and have key applications in the Wireless Sensor Network~(WSN) and Internet of Things~(IoT) world~\cite{Lopez2008}. While most works in this area deal with security \emph{protocols}~\cite{Pandey2010,Bruce2015}, implementing security mechanisms at the physical layer represents an interesting complement to those networking approaches~\cite{Bloch2011}, and has the potential to provide stronger (information-theoretic) secrecy properties~\cite{Shannon1949}. 

In the context of energy-constrained and green networking, the design of low-power systems and the use of renewable energy sources in network systems are prominent areas of investigation. In particular, the use of Energy Harvesting (EH) technologies as a way to prolong unattended operation of a network is becoming more and more appealing. However, despite these trends, security and privacy issues so far have been addressed mostly by neglecting low-power design principles (except possibly for some attempts at limiting the computation and processing costs and/or the number of messages needed to implement a secure protocol). In particular, the impact of power allocation policies and of system features related to energy harvesting has only been studied in some special cases~\cite{Alrajeh2013,Ng2014}. Since green aspects will play an increasingly large role in future networks, it is essential to bring low-power, energy-constrained and green considerations into this picture. In this paper, we try to partly fill this gap, studying how the use of energy harvesting affects the design and performance of physical layer security methods.

We consider an \emph{Energy Harvesting Device} (EHD) (i.e., a device with the capability of gathering energy from the environment~\cite{Gunduz2014}, e.g., through a solar panel or a rectenna) that sends data to a receiver over an insecure communication channel. The goal is to transmit data securely, i.e., in such a way that an adversary (or \emph{eavesdropper}) with access to the communication link is not able to gather useful information about the data sent. We study how the specific EH characteristics influence the achievable \emph{secrecy rate} (i.e., the information rate at which the EHD can reliably send data to the receiver while keeping it secret from the eavesdropper). Deciding whether the EHD should transmit or not, how much power should be transmitted or how to divide the power among the different sub-carriers is not obvious, and all these aspects need to be appropriately optimized. Moreover, while in the classic throughput optimization problem if the available resources were used improperly the corresponding penalty would be a performance reduction, in the secrecy optimization problem an improper use of the resources may imply not only a reduced transmission rate, but also a security loss, possibly making sensitive data accessible to a malicious party.

In the literature, many papers studied energy harvesting communication systems because of their ability to increase the network lifetime, provide self-sustainability and, ideally, allow perpetual operations~\cite{Ulukus2015}. \cite{Zhou2014} presented a survey on the several different environmental energy harvesting technologies for WSNs. Analytically, \cite{Lei2009} formulated the problem of maximizing the average value of the reported data using a node with a rechargeable battery. In~\cite{Sharma2010,Sharma2011}, Sharma \emph{et al.} studied heuristic delay-minimizing policies and sufficient stability conditions for a single EHD with a data queue. Ozel \emph{et al.} set up the offline throughput optimization problem from an information theoretic point of view in~\cite{Ozel2012b}, where they derived the information-theoretic capacity of the AWGN channel and presented two schemes that achieve such capacity (save-and-transmit and best-effort-transmit). In~\cite{Ozel2011}, the authors also modeled a battery-less system by a channel with state dependent amplitude constraints and causal information at the transmitter, and derived the capacity of this channel by making use of a result by Shannon. The throughput optimization problem with finite batteries in an EH system was studied in~\cite{Tutuncuoglu2012a,Michelusi2013}.

Security aspects have been widely studied in the WSN literature~\cite{Perrig2004,Lopez2008,Pandey2010}. Examples of relevant applications in a WSN/IoT context include health-care monitoring~\cite{AlAmeen2012,Agrawal2015}, where the sensitive data of patients may be exposed to a malicious party, or military use~\cite{Winkler2008,Jurisic2012}, where a WSN can be used for monitoring or tracking enemy forces. In particular, in addition to higher layers~\cite{Zin2014}, that are relatively insensitive to the physical characteristics of the wireless medium, physical layer can be used to strengthen the security of digital communication systems and improve already existing security measures. The basic idea behind the concept of physical layer secrecy is to exploit the randomness of the communication channel to limit the information that can be gathered by the eavesdropper at the signal level. Through channel coding techniques, it is possible to simultaneously allow the legitimate receiver to correctly decode a packet and prevent a potential third party malicious eavesdropper from decoding it and thus provide information-theoretic or \emph{unconditional} security. Differently from computational security methods, that are based on the limited computational capabilities of the adversary (as in a cryptographic system), unconditional security is considered the strongest notion of security~\cite{Maurer1993} because no limits on the adversary's computing power are assumed.
Perfect secrecy~\cite{Shannon1949} is achieved when there is zero mutual information between the information signal, $s$, and the signal received by the eavesdropper, $z$, i.e., $I(s;z) = 0$ and $z$ is useless when trying to determine $s$. In~\cite{Wyner1975}, Wyner showed that if the eavesdropper's channel is degraded with respect to the legitimate channel, then it is possible to exchange secure information at a non-zero rate while keeping the information leakage to the eavesdropper at a vanishing rate. This result was extended in~\cite{Csiszar1978} for non-degraded channels provided the eavesdropper channel is not less noisy than the legitimate channel. In~\cite{Wang2007}, the  secrecy capacity of fading channels in the presence of multiple eavesdroppers is studied. It was shown in~\cite{Gopala2008}~that in a fading scenario it is also possible to obtain a non-zero secure rate even if, on average, the eavesdropper's channel is better than the legitimate one. 
The authors also established the importance of variable rate coding (i.e., matching the code rate to the channel rate) in enabling secure communications. In~\cite{Oggier2011}, the authors compute the secrecy capacity of a MIMO wiretap channel with one receiver and one eavesdropper and an arbitrary number of antennas. A survey of physical layer security in modern networks is presented in~\cite{Mukherjee2014}.

The secrecy capacity paradigm in an energy harvesting communication system was studied in~\cite{Ozel2012,Ozel2012a}, where the authors considered the case of a batteryless transmitter and found the rate-equivocation region. 
\cite{Mukherjee2012} studied the deployment of an energy harvesting cooperative jammer to increase physical layer security.
In~\cite{Ng2014} the authors presented a resource allocation algorithm for a multiple-input single-output secrecy system for a communication system based on RF energy harvesting. Also~\cite{Zhang2014} studied how to efficiently allocate power over several sub-carriers in an EH system with secrecy constraints. In~\cite{Li2014} the authors employed a physical layer secrecy approach in a system with a transmitter that sends confidential messages to a receiver and transfers wireless energy to energy harvesting receivers. Our focus is substantially different from those: in the present paper we consider an EHD that harvests energy from an external, non-controllable and renewable energy source. Our goal is to maximize the achievable secrecy rate, i.e., to define how to correctly exploit the available (random) energy according to the device battery dynamics. 

Our main contribution lies in the definition of a new practical and challenging problem. As in~\cite{Ozel2012,Ozel2012a}, we investigate the physical layer secrecy in an EH system. However, differently from those papers, we explicitly consider the effects of a finite battery and we focus on finding the transmission strategy that maximizes the secrecy rate, namely the \emph{Optimal Secrecy Policy} (OSP). Since in a WSN the devices operate under the same conditions for long periods, the steady-state regime is generally reached, and thus we focus on the long-term optimization.
Similarly to~\cite{Michelusi2013,Biason2015e}, we set up an optimization problem based on a Markov Decision Process (MDP) approach but, unlike in those works, we focus on the security aspects, considering the presence of a malicious eavesdropper and a generic number of sub-carriers. Thus, even if the proposed analytical framework is similar to those provided in the literature, since additional dimensions are considered, the optimization process is more challenging and different considerations and insights are derived. In particular, we prove several properties of OSP and describe a technique to compute it by decomposing the problem into two steps. We specify how to allocate the power over the different sub-carriers and remark that a smart power splitting scheme is important to achieve high secrecy rates.
As~in \cite{Gopala2008}, we consider several degrees of knowledge of the channel state information, describing both variable and constant rate coding techniques and discussing how the achievable secrecy rate changes in these cases. However, unlike~\cite{Gopala2008}, we study an energy constrained system with $N$ parallel sub-carriers, and accordingly formulate and solve an optimization problem to determine the maximum secrecy rate.
Therefore, our paper considers aspects that either have not yet been considered or have been separately studied in the literature, and represents an advancement of the state of the art in the important areas of green networking and security, leading to novel insights about the interaction of many different system design aspects.

The paper is organized as follows. Section~\ref{sec:system_model} defines the system model we analyze and introduces the notion of secrecy rate. In Section~\ref{sec:optimization} we introduce the secrecy rate optimization problem. Section~\ref{sec:OSP_complete_CSI} describes how to find OSP and some of its properties with full CSI. In Section~\ref{sec:statistical_CSI} we study the case of imperfect CSI knowledge. Section~\ref{sec:numerical_evaluation} presents our numerical results. Finally, Section~\ref{sec:conclusions} concludes the paper.

\begin{figure}
  \centering
  \psfrag{transmitter}[c][c]{\footnotesize transmitter}
  \psfrag{receiver}[c][c]{\footnotesize receiver}
  \psfrag{eavesdropper}[c][c]{\footnotesize eavesdropper}
  \psfrag{harvested_energy}[c][c]{\footnotesize harvested energy}
  \psfrag{sigma_1}[c][c]{\footnotesize $\rho_1$}
  \psfrag{sigma_N}[c][c]{\footnotesize $\rho_N$}
  \psfrag{g_{1,k}}[c][c]{\footnotesize $\mathcal{g}_{1}$}
  \psfrag{h_{1,k}}[c][c]{\footnotesize $\mathcal{h}_{1}$}
  \psfrag{g_{N,k}}[c][c]{\footnotesize $\mathcal{g}_{N}$}
  \psfrag{h_{N,k}}[c][c]{\footnotesize $\mathcal{h}_{N}$}
  \includegraphics[width=0.9\columnwidth]{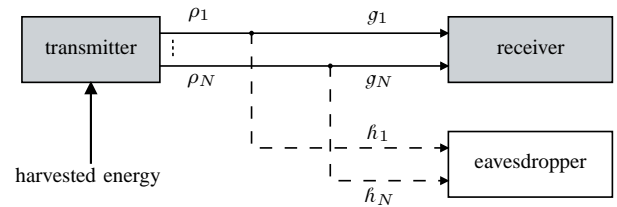}
  \caption{Block diagram of the system. $\boldsymbol{\mathcal{g}}$ and $\boldsymbol{\mathcal{h}}$ are the channel gains and $\boldsymbol{\rho}$ represents the power allocated over the $N$ sub-carriers.}
  \label{fig:block_diagram}
\end{figure}

\section{System Model and Secrecy Rate}\label{sec:system_model}

We consider an \emph{Energy Harvesting Device} (EHD) that simultaneously transmits data in a wide frequency band composed of $N$ narrow bands. The transmission power can be different for every sub-carrier. The transmission model can be described as a set of $N$ parallel Gaussian wiretap channels, affected by independent fading, as in \cite{Baldi2014}. The goal of the transmitter is to send data to the legitimate receiver with a positive secrecy rate in order to guarantee secure transmission. An eavesdropper attempts to intercept the transmitted data (see \figurename~\ref{fig:block_diagram} for the block diagram of the system model).

We initially assume that the EHD knows the Channel State Information (CSI) of all the sub-carriers toward the receiver and the eavesdropper instantaneously, and will relax this hypothesis in Section~\ref{sec:statistical_CSI}.
Time is divided into slots of equal duration $T$, chosen according to the channel coherence time, in order to guarantee constant channel gains in every slot.
The EHD is equipped with a battery of finite size $e_{\rm max}$ and in slot $k$ the device has $E^{(k)} \in \mathcal{E} \triangleq \{0,\ldots,e_{\rm max}\}$ energy quanta stored.\footnote{While in reality energy is a continuous quantity, we decide to adopt an approximate approach and discretize it in order to simplify the numerical optimization and apply the discrete MDP theory. However, we remark that it is always possible to use a finer quantization in order to improve the accuracy of the discrete approximation (which however implies higher complexity).} Knowledge of the state of charge is useful at the transmitter side only to determine when to schedule a transmission. The harvesting process is described through an energy quanta arrival process $\{B^{(k)}\}$, e.g., deterministic, Bernoulli or truncated geometric (for example, see~\cite{Gorlatova2013} for a characterization of the light energy). The average harvesting rate is $\bar{b}$, the maximum (minimum) number of energy quanta harvested per slot is $b_{\rm max}$ ($b_{\rm min}$), and a quantum harvested in slot $k$ can only be used in time slots $>k$. We assume that the device always has data to send and that the energy cost that the device sustains is mainly due to data transmission. Extensions to more general models are left for future work.

The channel gains in slot $k$ are $\boldsymbol{\mathcal{g}}^{(k)} = [\mathcal{g}_1^{(k)},\ldots,\mathcal{g}_N^{(k)}]$ and $\boldsymbol{\mathcal{h}}^{(k)} = [\mathcal{h}_1^{(k)},\ldots,\mathcal{h}_N^{(k)}]$ for the $N$ legitimate and eavesdropper sub-carriers, respectively. $\boldsymbol{\mathcal{g}}^{(k)}$ and $\boldsymbol{\mathcal{h}}^{(k)}$ can be interpreted as realizations of two jointly random vectors $\boldsymbol{G} = [G_1,\ldots,G_N]$ and $\boldsymbol{H} = [H_1,\ldots,H_N]$ (i.i.d. over time) with supports $\mathcal{G}$ and $\mathcal{H}$. We assume that the receiver has complete CSI of its channel in order to decode the received signal. Instead, the eavesdropper has knowledge on every aspect of the system (this is a reasonable worst-case assumption, as the transmission strategy should not rely on assuming the eavesdropper's ignorance of any state). Nevertheless, we should point out that, for a passive eavesdropper, knowledge of the main channel state is totally immaterial. In the following, when we refer to ``full'' or ``partial'' CSI, we always refer to the transmitter side.

\subsection{Secrecy Rates and Capacity}

We refer to the notions of \emph{secrecy rate} and \emph{secrecy capacity} as known in the physical layer secrecy literature \cite{Wyner1975,Bloch2011} and their ergodic counterparts in the fading scenario \cite{Liang2008}.
Specifically, we define an $(M,N,\ell)$ code for the parallel wiretap channel as consisting of: 1) a message set $\mathcal{S}$ with cardinality $M$, 2) a probabilistic encoder $f_\ell^{\rm enc}$ at the transmitter  that maps each message $s \in \mathcal{S} $ (realization of the r.v. $S$) to each $N\times \ell$ codeword $\B x \in \B{\C X}^\ell$, with  $\B{\C X} = \C X_1 \times \cdots \times \C X_N$ according to some conditional distribution $p_{\B X|S}(\B x|s)$, and 3) a (deterministic) decoder at the legitimate receiver that extracts $\hat{s}$ (realization of the r.v. $\hat{S}$) from the received message $\B y \in \B{\mathcal{Y}}^\ell$,  where $\B{\C Y} = \C Y_1 \times \cdots \times \C Y_N$ i.e., $f_\ell^{\rm dec} : \mathcal{Y^\ell} \rightarrow \mathcal{S}$.

The average error probability of an $(M,N,\ell)$ code is given by~
\begin{align}
  P_{\rm err}^\ell \triangleq \frac{1}{M}\sum_{s \in \mathcal{S}} \mathbb{P}\Big(\hat{S} \neq s | S = s\Big).
\end{align}

The equivocation rate at the eavesdropper is $R_e^\ell = (1/\ell) H(S | Z^\ell)$, i.e., the conditional entropy rate of the transmitted message given the eavesdropper's channel output $Z^\ell$. $R_e^\ell$ represents the level of ignorance on the target secret message at the eavesdropper. Perfect secrecy (unconditional security) would be obtained if $R_e^\ell = R^\ell$, where $R^\ell = (1/\ell) H(S)$ is the secret message rate. However, this is not possible in general with wiretap coding techniques, so we must settle for a weaker requirement, that holds asymptotically. Therefore, a \emph{secrecy rate} $R_s$ is said to be achievable if there exists a sequence of $(2^{\ell R_s},N,\ell)$ codes, $\ell = 1,2,\ldots,$ such that
\begin{align}\label{eq:unconditional_sec}
  &\lim_{\ell \rightarrow \infty} P_{\rm err}^\ell = 0, \qquad R_s \leq \lim_{\ell \rightarrow \infty} R_e^\ell 
\end{align}

\noindent and the secrecy capacity is defined as the supremum of the set of achievable secrecy rates. 

\subsection{Coding Strategy}\label{subsec:coding_strategy}

The transmitter coding strategy influences the secrecy rate. In particular, in this paper we consider \emph{constant} and \emph{variable} rate coding defined as follows (a construction procedure for these codes can be derived as explained in~\cite[Theorems~1 and~2]{Gopala2008}).

\textbf{Variable rate coding} consists in adapting the code rate to the main channel state. This can be accomplished by constructing a separate codeword $x$ for every realization of the channel, i.e., $x = x(\mbox{current channel})$. In this case, in every slot $k$ and on every sub-carrier $r = 1,\ldots,N$ the transmitter observes the channel and picks the symbols to be transmitted from the current codeword $x(\mathcal{g}_r^{(k)})$. We study the long-term regime and thus we consider the case of infinite length codewords.
	With variable rate coding, when the gain of the legitimate channel in a given sub-carrier is $\mathcal{g}$, the transmitter uses symbols from codewords at rate $\log(1 + \mathcal{g} \rho)$ (where $\rho$ is the transmission power, which will be the objective of our optimization). To achieve such a rate, it is required \emph{to use a codeword specifically designed for this channel}, i.e., $x(\mathcal{g})$. Then, if the eavesdropper's channel gain is $\mathcal{h} > \mathcal{g}$, thanks to the chosen coding rate, the mutual information between the transmitter and the eavesdropper is upper-bounded by $\log(1 + \mathcal{g} \rho)$. Instead, when $\mathcal{h} \leq \mathcal{g}$, the mutual information becomes $\log(1 + \mathcal{h} \rho)$ (Shannon's theorem). We can summarize the two previous cases as $\log(1+\min\{\mathcal{g},\mathcal{h}\}\rho)$.		
Therefore, even if $\mathcal{h} > \mathcal{g}$, the eavesdropper does not receive more information than the legitimate receiver (they both experience the same rate $\log(1 + \mathcal{g} \rho)$). In the long run, the average rate of the main channel and the information accumulated at the eavesdropper are~
\begin{align}
    \liminf_{K \rightarrow \infty} \frac{1}{K+1} \sum_{k = 0}^{K} \sum_{r = 1}^N \log(1 + \mathcal{g}_r^{(k)} \rho)
\end{align}

\noindent and~
\begin{align}
    \liminf_{K \rightarrow \infty} \frac{1}{K+1} \sum_{k = 0}^{K} \sum_{r = 1}^N \log(1 + \min\{\mathcal{g}_r^{(k)},\mathcal{h}_r^{(k)}\} \rho),
\end{align}

\noindent respectively. In this case, by constructing a code and the corresponding coding map, the long-term secrecy rate (amount of secret information that can be sent) is~
\begin{align}\label{eq:C_variable}
    \begin{split}
        \liminf_{K \rightarrow \infty} \frac{1}{K+1} \sum_{k = 0}^{K} \sum_{r = 1}^N  \Big(& \log(1 + \mathcal{g}_r^{(k)} \rho) \\
        &- \log(1 + \min\{\mathcal{g}_r^{(k)},\mathcal{h}_r^{(k)}\} \rho)\Big).
    \end{split}
\end{align}
	
\textbf{Constant rate coding} consists in keeping the code rate constant, regardless of the legitimate and eavesdropper's channel states. In this case, a single codeword $x$ is used in every fading condition. In every slot, the transmitter picks the symbols to be transmitter from the only available codeword $x$. In the long run, since we consider infinite length codewords, $x$ spans the entire fading statistic of the channel.
With constant rate coding, regardless of the current channel state, the transmitter uses codewords at a fixed rate $R_{\rm con}$ such that $R_{\rm con} \geq \log(1+\mathcal{g}\rho)$ for every $\mathcal{g}$ and $\rho$. In this case, if the current legitimate channel is $\mathcal{g}$, the mutual information between transmitter and receiver is upper bounded by Shannon's theorem as $\log(1+\mathcal{g}\rho)$. Similarly, the mutual information between transmitter and eavesdropper is given by $\log(1+\mathcal{h}\rho)$. The secrecy rate can be expressed as~
\begin{align}\label{eq:C_constant}
    {\left[\liminf_{K \rightarrow \infty} \! \frac{1}{K+1} \!  \sum_{k = 0}^{K} \sum_{r = 1}^N \! \Big(\! \log(1 + \mathcal{g}_r^{(k)} \rho) - \log(1 + \mathcal{h}_r^{(k)} \rho)\Big) \right]}^{{+}}\!\!\!,
\end{align}

\noindent where $[\cdot]^+ \triangleq \max\{0,\cdot\}$ is used to obtain a non-negative rate. Note that~\eqref{eq:C_constant} is lower than (or equal to)~\eqref{eq:C_variable}, i.e., higher secrecy is achieved with variable rate coding. However, its implementation is more difficult as the code rate has to be changed frequently according to the legitimate channel state.
\\

For simplicity, in the next we use $R_{\mathcal{g},\mathcal{h}}(\rho)$ to indicate the terms of the sum in~\eqref{eq:C_variable} if variable rate coding is considered, or~\eqref{eq:C_constant} in the constant rate coding case, i.e.,~
\begin{align}\label{eq:R_var_con}
	R_{\mathcal{g},\mathcal{h}}(\rho) &\triangleq \begin{cases}
	    \log(1 + \mathcal{g} \rho) - \log(1 + \min\{\mathcal{g},\mathcal{h}\} \rho), \quad &\mbox{var. rate}, \\
	    \log(1 + \mathcal{g} \rho) - \log(1 + \mathcal{h} \rho), \quad &\mbox{con. rate}. \\
	\end{cases}
\end{align}

\noindent $c(\boldsymbol{\rho},\boldsymbol{\mathcal{g}},\boldsymbol{\mathcal{h}})$ is the generalization with a generic number of sub-carriers $N$:~
\begin{align}
  &c(\boldsymbol{\rho},\boldsymbol{\mathcal{g}},\boldsymbol{\mathcal{h}}) = \sum_{r = 1}^{N} R_{\mathcal{g}_r,\mathcal{h}_r}(\rho_r), \label{eq:c_rho_c_r}
\end{align} 

\noindent and $\rho^{\rm tot}$ is the corresponding total transmission power, defined as~
\begin{align}
  \rho^{\rm tot} &\triangleq \mathbf{1}_N^T \boldsymbol{\rho}. \label{eq:omega_equal_sum_sigma}
\end{align}

\noindent The value of $c(\boldsymbol{\rho},\boldsymbol{\mathcal{g}},\boldsymbol{\mathcal{h}})$ depends on the choice of the power allocation over the several sub-carriers, $\boldsymbol{\rho} \triangleq [\rho_1,\ldots,\rho_N]^T$, the channel conditions $\boldsymbol{\mathcal{g}}$ and $\boldsymbol{\mathcal{h}}$, and the coding rate strategy. $\mathbf{1}_N$ is a column vector consisting of $N$ ones. In the general case, the choice of $\boldsymbol{\rho}$ that maximizes the secrecy rate, among those satisfying~\eqref{eq:omega_equal_sum_sigma}, will in turn depend upon the channel conditions $\boldsymbol{\mathcal{g}}$ and $\boldsymbol{\mathcal{h}}$.

\section{Optimization Problem}\label{sec:optimization}

The system state $\mathbf{S}^{(k)}$ in time slot $k$ is defined by the $(2N+1)$-tuple $(E^{(k)},\boldsymbol{\mathcal{g}}^{(k)},\boldsymbol{\mathcal{h}}^{(k)})$. 
A policy $\mu$ is a set of rules that, given the state of the system, specifies the power allocation over the $N$ sub-carriers.

In the long run, the average \emph{secrecy rate} under a policy~$\mu$ is given by the average undiscounted reward~$C_{\mu}$~
\begin{align}
    C_{\mu}(E^{(0)}) \triangleq \left[\liminf_{K \rightarrow \infty} \frac{1}{K+1} \sum_{k = 0}^{K} c(\boldsymbol{\Sigma}^{(k)},\boldsymbol{\mathcal{g}}^{(k)},\boldsymbol{\mathcal{h}}^{(k)})\right]^+\!\!\!\!\!, \label{eq:C_omega}
\end{align}

\noindent where $c(\cdot,\cdot,\cdot)$ is the instantaneous partial contribution defined in~\eqref{eq:c_rho_c_r}, $\boldsymbol{\Sigma}^{(k)}$ is the power allocation vector defined by the policy\footnote{Given a temporal sequence of energy arrivals and channel states, the policy $\mu$ can be applied to obtain the power allocation vector $\boldsymbol{\Sigma}^{(k)}$. In this case we use a deterministic policy for presentation simplicity, and prove later that this choice is optimal.} and $E^{(0)}$ is the energy in the initial time slot. A secure communication can be performed if $C_{\mu}(E^{(0)}) > 0$. \eqref{eq:C_omega}~is a generalization of~\eqref{eq:C_variable} and~\eqref{eq:C_constant} for $N$ sub-carriers and a dynamic transmission power.

The battery evolution is as follows~
\begin{align}\label{eq:Ep1_E}
    E^{(k+1)} = \min\left\{E^{(k)} - \sum_{r=1}^N \Sigma_r^{(k)} + B^{(k)}, e_{\rm max}\right\},
\end{align}

\noindent where $\Sigma_r^{(k)}$ is the $r^{th}$ component of the vector $\boldsymbol{\Sigma}^{(k)}$, and the $\min$ is used to account for the finite battery. Note that $\boldsymbol{\Sigma}^{(k)}$ must satisfy $\sum_{r=1}^N \Sigma_r^{(k)} \leq E^{(k)},\ \forall k$ and $\Sigma_r^{(k)} \geq 0,\ \forall k,\ \forall r$. Thus, Problem~\eqref{eq:C_omega} is implicitly influenced by the evolution of $E^{(k)}$ because of $\boldsymbol{\Sigma}^{(k)}$.

Our aim is to solve the following maximization problem~
\begin{align}
    \mu^\star = &\ \argmax{\mu} C_\mu(E^{(0)}).\label{eq:max_problem}
\end{align}

A policy that solves \eqref{eq:max_problem} is an \emph{Optimal Secrecy Policy} (OSP). In the next subsection we explain in more detail the optimization variables and the constraints of the above problem.

\subsection{Markov Decision Process Formulation}
Since we consider a \emph{long-term} optimization, we recast the problem using a Markov Decision Process~(MDP) formulation. In particular, we model our system by a Markov Chain (MC) with a finite number of states. For every MC state $(e,\boldsymbol{\mathcal{g}},\boldsymbol{\mathcal{h}})$, a \emph{power allocation policy} $\mu$ is the set of rules~
\begin{align}\label{eq:mu}
    \mu = \{\mu(\cdot;e,\boldsymbol{\mathcal{g}},\boldsymbol{\mathcal{h}}),\ \forall e \in \mathcal{E},\ \forall \boldsymbol{\mathcal{g}} \in \mathcal{G}, \ \boldsymbol{\mathcal{h}} \in \mathcal{H}\},
\end{align}

\noindent where $\mu(\cdot;e,\boldsymbol{\mathcal{g}},\boldsymbol{\mathcal{h}})$ is the conditional distribution (pmf) of the power allocation vector defined as follows~
\begin{align}
    &\mu(\boldsymbol{\rho};e,\boldsymbol{\mathcal{g}},\boldsymbol{\mathcal{h}}) \triangleq \mathbb{P}\left( \substack{\mbox{using a power} \\ \mbox{splitting vector $\boldsymbol{\rho}$}} \ \!\big| e,\! \boldsymbol{G} \!=\! \boldsymbol{\mathcal{g}},\!\boldsymbol{H} \!=\! \boldsymbol{\mathcal{h}} \right)\!,
\end{align}

\noindent and, for every $\boldsymbol{\mathcal{g}}$, $\boldsymbol{\mathcal{h}}$, is subject to~
\begin{subequations}
\begin{align}
    &\sum_{\mathclap{\boldsymbol{\rho} \in \mathcal{P}_{\scriptscriptstyle \leq}(e)}}\mu(\boldsymbol{\rho};e,\boldsymbol{\mathcal{g}},\boldsymbol{\mathcal{h}}) = 1, \\
    &\mu(\boldsymbol{\rho};e,\boldsymbol{\mathcal{g}},\boldsymbol{\mathcal{h}}) \geq 0, \qquad \forall \boldsymbol{\rho} \in \mathcal{P}_{\scriptscriptstyle \leq}(e), \\
    & \mathcal{P}_{\scriptscriptstyle \leq}(e) \triangleq \left\{\boldsymbol{\rho}\ : \ \boldsymbol{\rho} \succeq 0 \ \cap \ \rho^{\rm tot} \triangleq \mathbf{1}_N^T \boldsymbol{\rho} \leq e \right\}.
\end{align}
\label{eq:mu_constraints}
\end{subequations}

\noindent $\mathcal{P}_{\scriptscriptstyle \leq}(e)$ is the set of all feasible vectors $\boldsymbol{\rho}$ when the energy level is $e$. The reward function becomes~
\begin{align}  
      C_{\mu}(E^{(0)}) =& \sum_{e \in \mathcal{E}}\pi_{\mu}(e|E^{(0)})  \label{eq:C_mu_E^{(0)}}\\
      & \times \!\! \int_{\mathcal{G}\times \mathcal{H}} \underbrace{\sum_{\boldsymbol{\rho} \in \mathcal{P}_{\scriptscriptstyle \leq}(e)} \!\! c(\boldsymbol{\rho},\boldsymbol{\mathcal{g}},\boldsymbol{\mathcal{h}}) \mu(\boldsymbol{\rho};e,\boldsymbol{\mathcal{g}},\boldsymbol{\mathcal{h}})}_{\mathclap{\mbox{\footnotesize secrecy rate given the MC state } (e,\boldsymbol{\mathcal{g}},\boldsymbol{\mathcal{h}})}} \ \mbox{d}F(\boldsymbol{\mathcal{g}},\boldsymbol{\mathcal{h}}), \nonumber
\end{align}

\noindent where $\pi_{\mu}(e|E^{(0)}) \in [0,1]$ is the steady-state probability of having $e$ energy quanta stored starting from state $E^{(0)}$ under a policy $\mu$ and $F (\boldsymbol{\mathcal{g}},\boldsymbol{\mathcal{h}})$ is the joint cumulative distribution function of~$\boldsymbol{G}$ and~$\boldsymbol{H}$. $\pi_\mu(e|E^{(0)})$ summarizes the battery evolution and is evaluated according to~\eqref{eq:Ep1_E}. The optimization variables in Problem~\eqref{eq:max_problem} are the pmfs $\mu(\cdot;e,\boldsymbol{\mathcal{g}},\boldsymbol{\mathcal{h}})$. Also, it can be shown (see Section~\ref{sec:unichain}) that an OSP which admits steady-state distribution always exists. Therefore, without loss of optimality, we decided to restrict our study to the class of policies with steady-state distribution. For these policies, since we focus on the average long-term optimization, \eqref{eq:C_mu_E^{(0)}}~is equivalent to~\eqref{eq:C_omega}.

It is possible to separate $\mu$ into the product of a \emph{transmit power policy}, which specifies the conditional distribution of the total transmission power given the current state, namely $\gamma_\mu(\rho^{\rm tot};e,\boldsymbol{\mathcal{g}},\boldsymbol{\mathcal{h}})$, and the conditional distribution of the power allocation given the total transmission power and the current state, namely $\phi_\mu(\boldsymbol{\rho};\rho^{\rm tot},e,\boldsymbol{\mathcal{g}},\boldsymbol{\mathcal{h}})$:~
\begin{align}
    \mu(\boldsymbol{\rho};e,\boldsymbol{\mathcal{g}},\boldsymbol{\mathcal{h}}) = \phi_\mu(\boldsymbol{\rho};\rho^{\rm tot},e,\boldsymbol{\mathcal{g}},\boldsymbol{\mathcal{h}})\gamma_\mu(\rho^{\rm tot};e,\boldsymbol{\mathcal{g}},\boldsymbol{\mathcal{h}}). \label{eq:mu_decomposition}
\end{align}

The above expression will be useful to decompose the problem into two steps in Theorem~\ref{thm:max_C_Omega}.

We highlight that $\mu$ performs a \emph{power control} mechanism, i.e., it specifies how much power is used in every MC state but, in addition to power control, also the code rate can be changed according to Section~\ref{subsec:coding_strategy}.

\subsection{Finite Model}\label{sec:finite_model}

In the previous subsection, we assumed that the policy can be defined for every possible value of the channel gains. This can be done by simple enumeration if $|\mathcal{G}| <\infty$ and $|\mathcal{H}| < \infty$. However, the channel gains may be continuous variables in the general case. Instead of defining a policy for a continuously infinite set of values, we want to find a set of points where the policy can be computed and optimized efficiently. The following approach can be followed. Consider the random variable $G_1$ (for the others the reasoning is similar). We discretize the support of $G_1$ in $n$ intervals with an \emph{equally likely} strategy ($\mathbb{P}(G_1 \in [p_i,p_{i+1})) = 1/n$, $i = 1,\ldots,n$). Then, we specify the policy in the centroid of every interval. If the number of intervals $n$ is sufficiently large, the approximation is very close to the continuous case.

\begin{remark}\label{remark:non_null}
    Since we consider a discrete channel, we focus without loss of generality on channel conditions with non-zero probability, i.e., $\mathbb{P}(\boldsymbol{G} = \boldsymbol{\mathcal{g}}, \boldsymbol{H} = \boldsymbol{\mathcal{h}})>0$, $\forall \boldsymbol{\mathcal{g}} \in \mathcal{G}, \boldsymbol{\mathcal{h}} \in \mathcal{H}$.
\end{remark}

\section{Optimal Secrecy Policy with Complete CSI}\label{sec:OSP_complete_CSI}

In this section we study the case when the transmitter has perfect CSI knowledge, and introduce a technique to compute OSP and some of its properties. All our results are useful to simplify the numerical evaluation. In particular: 1) we prove that there exists a deterministic OSP (Theorem~\ref{thm:OSP_deterministic}); 2) we propose a technique to derive a unichain OSP (Section~\ref{sec:unichain}); 3) we decompose the optimization process in two steps (Theorem~\ref{thm:max_C_Omega}); and 4) we show that the transmission power increases (decreases) with the channel gain of the legitimate receiver's (eavesdropper's) sub-carriers (Theorem~\ref{thm:OSP_increasing_decreasing}).

\begin{thm} \label{thm:OSP_deterministic}
    There exists a deterministic OSP, i.e., an optimal secrecy policy in which, for every MC state $(e,\boldsymbol{\mathcal{g}},\boldsymbol{\mathcal{h}})$~
    \begin{align} \label{eq:OSP_deterministic}
        \mu^\star(\boldsymbol{\rho}; e, \boldsymbol{\mathcal{g}},\boldsymbol{\mathcal{h}}) = 
        \begin{cases}
        1, \qquad &\mbox{if } \boldsymbol{\rho} = \boldsymbol{\rho}_{e,\boldsymbol{\mathcal{g}},\boldsymbol{\mathcal{h}}}^\star, \\
        0, \qquad &\mbox{otherwise},
        \end{cases}
    \end{align}
    
    \noindent for some $\boldsymbol{\rho}_{e,\boldsymbol{\mathcal{g}},\boldsymbol{\mathcal{h}}}^\star$ depending upon the current MC state in general.
    \begin{proof}
        See Appendix~\ref{proof:OSP_deterministic}.
    \end{proof}
\end{thm}

By exploiting Equation~\eqref{eq:mu_decomposition}, it also follows that $\exists \rho_{e,\boldsymbol{\mathcal{g}},\boldsymbol{\mathcal{h}}} ^{{\rm tot}^{\scriptstyle \star}}$ such that the transmit power policy $\gamma_\mu$ defined in~\eqref{eq:mu_decomposition} satisfies~
\begin{align}
    \gamma_\mu(\rho^{\rm tot}; e, \boldsymbol{\mathcal{g}},\boldsymbol{\mathcal{h}}) = 
    \begin{cases}
        1, \qquad &\mbox{if } \rho^{\rm tot} = \rho_{e,\boldsymbol{\mathcal{g}},\boldsymbol{\mathcal{h}}} ^{{\rm tot}^{\scriptstyle \star}}, \\
        0, \qquad &\mbox{otherwise}.
    \end{cases}
\end{align}

\begin{definition}[Deterministic Policy]
Since a deterministic OSP always exists, we only need to study deterministic policies, thus $\mu$ can be redefined as~
\begin{align} \label{eq:mu_det}
    \mu = \{\boldsymbol{\rho}_{e,\boldsymbol{\mathcal{g}},\boldsymbol{\mathcal{h}}} \in \mathcal{P}_{\scriptscriptstyle \leq}(e),\ \forall e \in \mathcal{E},\ \forall \boldsymbol{\mathcal{g}} \in \mathcal{G}, \ \boldsymbol{\mathcal{h}} \in \mathcal{H}\}.
\end{align}

\noindent $\boldsymbol{\rho}_{e,\boldsymbol{\mathcal{g}},\boldsymbol{\mathcal{h}}} = [\rho_{1;e,\boldsymbol{\mathcal{g}},\boldsymbol{\mathcal{h}}},\ldots,\rho_{N;e,\boldsymbol{\mathcal{g}},\boldsymbol{\mathcal{h}}}]$ characterizes the transmission powers on different sub-carriers in state $(e,\boldsymbol{\mathcal{g}},\boldsymbol{\mathcal{h}})$.
\end{definition}

We also introduce the sub-policy $\mu^{\rm tot}$ as~
\begin{align} \label{eq:mu_tot_det}
    \mu^{\rm tot} = \{\rho^{\rm tot}_{e,\boldsymbol{\mathcal{g}},\boldsymbol{\mathcal{h}}},\ \forall e \in \mathcal{E},\ \forall \boldsymbol{\mathcal{g}} \in \mathcal{G}, \ \boldsymbol{\mathcal{h}} \in \mathcal{H}\},
\end{align}

\noindent which accounts for the total transmission powers only. $\mu^{\rm tot}$ and $\mu$ are \emph{consistent} if the sum of the elements of $\boldsymbol{\rho}_{e,\boldsymbol{\mathcal{g}},\boldsymbol{\mathcal{h}}}$ in $\mu$ is equal to ${\rho^{\rm tot}_{e,\boldsymbol{\mathcal{g}},\boldsymbol{\mathcal{h}}}}$ in $\mu^{\rm tot}$, $\forall e\in \mathcal{E}, \boldsymbol{\mathcal{g}} \in \mathcal{G}, \boldsymbol{\mathcal{h}} \in \mathcal{H}$.

The deterministic property is particularly useful to simplify the numerical evaluation because a policy needs to define only a scalar value for every state of the system and not a probability distribution.

\subsection{Unichain Policies}\label{sec:unichain}

We restrict our study to the class of \emph{unichain} policies, i.e., those that induce a unichain MC (i.e., a MC with a single recurrent class). This is useful in order to apply the standard optimization algorithms in the next section.

Some sufficient conditions to obtain a unichain policy are presented in the following proposition (in this subsection we use deterministic policies for presentation simplicity, but the results can be easily extended).

\begin{propos}\label{propos:irreducible}
    If a policy satisfies one of the following conditions, then it is unichain. If it satisfies both conditions, the policy induces an irreducible, positive recurrent MC.~
    \begin{enumerate}
        \item For every $e \in \mathcal{E} \backslash \{e_{\rm max}\}$ there exists a pair $(\boldsymbol{\mathcal{g}}',\boldsymbol{\mathcal{h}}')$ such that $\rho^{\rm tot}_{e,\boldsymbol{\mathcal{g}}',\boldsymbol{\mathcal{h}}'} < b_{\rm max}$ (maximum number of energy arrivals).
        \item For every $e \in \mathcal{E} \backslash \{0\}$ there exists a pair $(\boldsymbol{\mathcal{g}}'',\boldsymbol{\mathcal{h}}'')$ such that $\rho^{\rm tot}_{e,\boldsymbol{\mathcal{g}}'',\boldsymbol{\mathcal{h}}''} > b_{\rm min}$.
    \end{enumerate}
    \begin{proof}
        See Appendix~\ref{proof:irreducible}.
    \end{proof}
\end{propos}

In practice, the first and second points ensure that there is a positive probability that the battery moves from level $e$ to higher and lower energy levels, respectively. When they are both verified, no transient state can exist, and the MC is irreducible. 

When at least one point of Proposition~\ref{propos:irreducible} is satisfied, the corresponding policy is guaranteed to be unichain. However, in general, these conditions may not be satisfied and a policy may not be unichain. In addition, there may exist more than one policy with the same maximum achievable secrecy rate (the highest secrecy rate among $C_\mu(0),\ldots,C_\mu(e_{\rm max})$). Some of these are unichain, whereas others are not. Consider the following example to justify these claims.

\begin{example}\label{ex:no_unichain}
    We want to show a case in which 1) multiple policies with the same maximum reward exist and 2) some of them are not unichain.
    
    Assume that the harvesting process is deterministic and equal to $b_{\rm max} < e_{\rm max}/2$, $N = 1$, and the channel is constant $\mathcal{g}_1 > \mathcal{h}_1$. Consider the following policies~
    \begin{align*}
        \mu_1 =&\ \{\rho_{1;e,\mathcal{g}_1,\mathcal{h}_1} = \min\{e,b_{\rm max}\},\ \forall e,\ \forall \mathcal{g}_1, \ \mathcal{h}_1\}, \\
        \mu_2 =&\ \begin{Bmatrix*}[l]
            \rho_{1;e,\mathcal{g}_1,\mathcal{h}_1} = 2b_{\rm max}, \qquad & e = e_{\rm max},\ \forall \mathcal{g}_1, \ \mathcal{h}_1  \\
            \rho_{1;e,\mathcal{g}_1,\mathcal{h}_1} = b_{\rm max}, \qquad & e = b_{\rm max},\ \forall \mathcal{g}_1, \ \mathcal{h}_1 \\
            \rho_{1;e,\mathcal{g}_1,\mathcal{h}_1} = 0, \qquad & \mbox{otherwise}
        \end{Bmatrix*}.
    \end{align*}
   
    $\mu_1$ is a unichain policy (the recurrent class is the battery level $\{b_{\rm max}\}$) that provides a long-term secrecy rate $c(b_{\rm max},\mathcal{g}_1,\mathcal{h}_1)$. Instead, $\mu_2$ is not unichain (the two recurrent classes are  $\{b_{\rm max}\}$ and $\{e_{\rm max}-b_{\rm max},e_{\rm max}\}$) and its long-term secrecy rate depends upon the initial state (it can be $c(b_{\rm max},\mathcal{g}_1,\mathcal{h}_1)$ or $0.5c(2b_{\rm max},\mathcal{g}_1,\mathcal{h}_1)$). Also, note that because of the concavity of Equation~\eqref{eq:c_rho_c_r}, $c(b_{\rm max},\mathcal{g}_1,\mathcal{h}_1) > 0.5c(2b_{\rm max},\mathcal{g}_1,\mathcal{h}_1)$. Therefore, there exist more than one policy with the same maximum achievable reward $c(b_{\rm max},\mathcal{g}_1,\mathcal{h}_1)$. Moreover, in $\mu_2$, there are two recurrent classes, and thus it is not unichain.
\end{example}

This example shows that the long-term secrecy rate for a non-unichain policy may depend upon the starting state. Also, it shows that in general there may exist different policies, unichain and not unichain, with the same maximum achievable secrecy rate. The following proposition establishes that there is no loss in generality in considering only unichain policies.

\begin{propos}\label{propos:unichain}
    Given a generic policy, it is always possible to derive another policy which is unichain and attains the same maximum achievable secrecy rate as the original policy, regardless of the initial state.
    \begin{proof}
        We provide a constructive proof in Appendix~\ref{app:derive_unichain}.
    \end{proof}
\end{propos}

In the rest of the paper we always refer to unichain policies, for which $C_\mu(E^{(0)})$ is independent of $E^{(0)}$~\cite{Levin2009}. In particular, Proposition~\ref{propos:unichain} holds for the optimal secrecy policies, i.e., there always exists a unichain OSP, and therefore we will focus on unichain policies with no loss in optimality. Note that, since we consider a finite MC (we discretized both the battery level and the channel gains), a unichain policy always implies the existence of a steady-state distribution as in Equation~\eqref{eq:C_mu_E^{(0)}}.

\subsection{Computation of OSP}

We now want to simplify the expression of $C_\mu$ by exploiting the results we have found so far.
If $\mu$ and $\mu^{\rm tot}$ are consistent, the long-term secrecy function $C_\mu$ can be rewritten as~
\begin{align}
    C_\mu = \sum_{e \in \mathcal{E}} \pi_{\mu^{\rm tot}}(e)\int_{\mathcal{G}\times \mathcal{H}}& c(\overbrace{\boldsymbol{\rho}_{e,\boldsymbol{\mathcal{g}},\boldsymbol{\mathcal{h}}}}^{\mathclap{\mbox{\footnotesize specified by $\mu$}}},\boldsymbol{\mathcal{g}},\boldsymbol{\mathcal{h}}) \ \mbox{d}F(\boldsymbol{\mathcal{g}},\boldsymbol{\mathcal{h}}). \label{eq:C_fading_simplified}
\end{align}

An interesting fact is that the steady-state probability $\pi_{\mu^{\rm tot}}(e)$ depends upon the sub-policy $\mu^{\rm tot}$ only. This is because $\pi_{\mu^{\rm tot}}(e)$ describes the battery energy evolution, that depends only upon the total energy consumption in a slot, not upon the particular power splitting scheme. This result leads to the following theorem.

\begin{thm}\label{thm:max_C_Omega}
  The maximization of $C_{\mu}$ can be decomposed into two steps:
  \begin{enumerate}
    \item fix a value $x$ and the channel gain vectors $\boldsymbol{\mathcal{g}}$, $\boldsymbol{\mathcal{h}}$ and find the optimal power splitting choice~
    \begin{subequations}
    \begin{flalign}
      &\boldsymbol{\rho}^\star = \argmax{\boldsymbol{\rho}} c(\boldsymbol{\rho},\boldsymbol{\mathcal{g}},\boldsymbol{\mathcal{h}}), && \\
      \mbox{s.t.:} \quad & \boldsymbol{\rho} \in \mathcal{P}_{\scriptscriptstyle =}(x) \triangleq \left\{\boldsymbol{\rho}\ : \ \boldsymbol{\rho} \succeq 0,\ x = \mathbf{1}_N^T \boldsymbol{\rho} \right\}; &&\label{eq:sigma_star2}
    \end{flalign}
    \label{eq:sigma_star}
    \end{subequations}
    
    \item maximize $C_\mu$ by considering only $\mu^{\rm tot}$~    
    \begin{subequations}
    \begin{flalign}
        &\mu^{{\rm tot}^{\scriptstyle \star}} = \argmax{\mu^{\rm tot}} C_{\mu}, &&\\
        \mbox{s.t.:} \quad 
            &\mu^{\rm tot} \mbox{ and } \mu \mbox{ are consistent,} &&\\
            &\boldsymbol{\rho}_{e,\boldsymbol{\mathcal{g}},\boldsymbol{\mathcal{h}}} \mbox{ solves~\eqref{eq:sigma_star} with } x = {\rho^{\rm tot}_{e,\boldsymbol{\mathcal{g}},\boldsymbol{\mathcal{h}}}},\atop \forall e \in \mathcal{E}, \forall \boldsymbol{\mathcal{g}} \in \mathcal{G},\ \forall \boldsymbol{\mathcal{h}} \in \mathcal{H}. &&
    \end{flalign}
    \end{subequations}
  \end{enumerate}
  
  The optimal $\mu^\star$ can be found by fixing $\rho^{{\rm tot}^{\scriptstyle \star}}$ according to point 2) and choosing $\boldsymbol{\rho}$ with the optimal power splitting choice of point 1).
  
  \begin{proof}
      See Appendix~\ref{proof:max_C_Omega}.
  \end{proof}
\end{thm}

The \emph{optimal power splitting choice} $\boldsymbol{\rho}^\star$ that solves~\eqref{eq:sigma_star} can be found with a Lagrangian approach (for further details, see Theorem~1 and Equation~(7) in~\cite{Gopala2008}):~
\begin{align}
  &\rho^\star_r = \left[\sqrt{\frac{\alpha_r^2}{4} + \frac{\alpha_r}{\eta}} - \frac{\beta_r}{2}\right]^+,\label{eq:tau_r} \\
  &\alpha_r \triangleq \frac{1}{\mathcal{h}_r}-\frac{1}{\mathcal{g}_r},\qquad \beta_r \triangleq \frac{1}{\mathcal{h}_r}+\frac{1}{\mathcal{g}_r}, \label{eq:alpha_beta}
\end{align}

\noindent where $\eta$ is a parameter used to satisfy $x = \sum_{r=1}^N \rho_r^\star$.
In the remainder of the paper we assume that this optimal power splitting choice is used, unless otherwise stated. 
We highlight that OSP yields $\rho_r^\star = 0$ if $\mathcal{g}_r \leq \mathcal{h}_r$, which implies that the achievable secrecy rate with complete CSI is independent of the coding scheme (the two expressions in Equation~\eqref{eq:R_var_con} coincide).

To solve Step 2) instead, the Optimal Secrecy Policy can be found numerically via dynamic programming techniques, e.g., using the Policy Iteration Algorithm~(PIA)~\cite{Bertsekas2005}.\footnote{A key assumption of PIA is that, at every algorithm step, a \emph{unichain} policy is produced. In order to satisfy this condition, we apply the technique of Appendix~\ref{app:derive_unichain}.} 
PIA alternates between a value determination phase, in which the current policy is evaluated, and a policy improvement phase, in which an attempt is made at improving the current policy. Policy improvement and evaluation can be performed in $\mathcal{O}((e_{\rm max})^3 n^{2N})$ and $\mathcal{O}((e_{\rm max})^3)$ arithmetic operations, respectively, where $\mathcal{O}(\cdot)$ is the standard asymptotic notation. This result is derived as follows. For every state of the system ($e_{\rm max} \times n^N \times n^N$), the policy improvement step requires to find the best transmission power (which is $\mathcal{O}(e_{\rm max})$) to reach every other battery level ($e_{\rm max}$). Instead, the $\mathcal{O}((e_{\rm max})^3)$ performance of the policy evaluation step is due to a matrix inversion cost (which can be reduced to $\mathcal{O}((e_{\rm max})^{2.373})$ using Coppersmith-Winograd like algorithms).
The previous two steps are performed iteratively until the optimal policy is found, which, in general, requires few iterations ($< 10$). Therefore, PIA has a polynomial complexity in the number of states of the system.

Note that Theorem~\ref{thm:max_C_Omega} with~\eqref{eq:tau_r}-\eqref{eq:alpha_beta} decompose the optimization into two steps. Therefore, the numerical evaluation only requires to study the two points separately instead of performing a (more computationally intensive) bi-dimensional optimization.


We also remark the following.
\begin{lemma}
    By restricting the study to the unichain policies constructed as in Appendix~\ref{app:derive_unichain}, OSP is uniquely determined.
    \begin{proof}
        In all the transient states, by construction (Appendix~\ref{app:derive_unichain}), we have $\rho_{e,\boldsymbol{\mathcal{g}},\boldsymbol{\mathcal{h}}} ^{{\rm tot}^{\scriptstyle \star}} = 0$. For the recurrent states, thanks to~\cite[Vol. II, Sec. 4]{Bertsekas2005}, we know that $\rho_{e,\boldsymbol{\mathcal{g}},\boldsymbol{\mathcal{h}}} ^{{\rm tot}^{\scriptstyle \star}}$ is uniquely determined.
    \end{proof}
\end{lemma}

\subsection{Properties}
We now derive a property that is useful to understand when the transmission power increases or decreases.
\begin{propos} \label{propos:D}
Consider two channel states $\boldsymbol{\mathcal{g}}'$, $\boldsymbol{\mathcal{h}}'$ and $\boldsymbol{\mathcal{g}}''$, $\boldsymbol{\mathcal{h}}''$ and define
    \begin{align}
        &D(\rho^{\rm tot};\boldsymbol{\mathcal{g}}',\boldsymbol{\mathcal{h}}';\boldsymbol{\mathcal{g}}'',\boldsymbol{\mathcal{h}}'') \\
        &\triangleq \frac{\partial}{\partial \rho^{\rm tot}}\Big(c(\boldsymbol{\rho}_{e,\boldsymbol{\mathcal{g}}'',\boldsymbol{\mathcal{h}}''}^\star,\boldsymbol{\mathcal{g}}'',\boldsymbol{\mathcal{h}}'') - c(\boldsymbol{\rho}_{e,\boldsymbol{\mathcal{g}}',\boldsymbol{\mathcal{h}}'}^\star,\boldsymbol{\mathcal{g}}',\boldsymbol{\mathcal{h}}')\Big), \nonumber
    \end{align}
    \noindent where $\boldsymbol{\rho}_{e,\boldsymbol{\mathcal{g}}'',\boldsymbol{\mathcal{h}}''}^\star$ and $\boldsymbol{\rho}_{e,\boldsymbol{\mathcal{g}}',\boldsymbol{\mathcal{h}}'}^\star$ are defined as the solutions\footnote{Note that $\boldsymbol{\rho}_{e,\boldsymbol{\mathcal{g}}'',\boldsymbol{\mathcal{h}}''}^\star$ and $\boldsymbol{\rho}_{e,\boldsymbol{\mathcal{g}}',\boldsymbol{\mathcal{h}}'}^\star$ depend upon $\rho^{\rm tot}$.} of Problem~\eqref{eq:sigma_star} with $x = \rho^{\rm tot}$.

    OSP has the following trend~
    \begin{itemize}
        \item if $D(\rho^{\rm tot};\boldsymbol{\mathcal{g}}',\boldsymbol{\mathcal{h}}';\boldsymbol{\mathcal{g}}'',\boldsymbol{\mathcal{h}}'') \geq 0,\ \forall \rho^{\rm tot}$, then $\rho^{{\rm tot}^{\scriptstyle \star}}_{e,\boldsymbol{\mathcal{g}}'',\boldsymbol{\mathcal{h}}''} \geq \rho^{{\rm tot}^{\scriptstyle \star}}_{e,\boldsymbol{\mathcal{g}}',\boldsymbol{\mathcal{h}}'}$;
        \item if $D(\rho^{\rm tot};\boldsymbol{\mathcal{g}}',\boldsymbol{\mathcal{h}}';\boldsymbol{\mathcal{g}}'',\boldsymbol{\mathcal{h}}'') \leq 0,\ \forall \rho^{\rm tot}$, then $\rho^{{\rm tot}^{\scriptstyle \star}}_{e,\boldsymbol{\mathcal{g}}'',\boldsymbol{\mathcal{h}}''} \leq \rho^{{\rm tot}^{\scriptstyle \star}}_{e,\boldsymbol{\mathcal{g}}',\boldsymbol{\mathcal{h}}'}$.
    \end{itemize}
    
    \begin{proof}
        See Appendix~\ref{proof:D}.
    \end{proof}
\end{propos}

In practice, it is better to use more energy in the directions where the function $c(\cdot,\cdot,\cdot)$ increases. A consequence of the previous proposition is derived in the following theorem.

\begin{thm}\label{thm:OSP_increasing_decreasing}
    Consider $N = 1$. The transmission power of OSP is non-decreasing with $\mathcal{g}$ and non-increasing with $\mathcal{h}$ (we omit the ``$1$'' subscripts). Formally~
    \begin{itemize}
        \item if $\mathcal{g}'' \geq \mathcal{g}'$, then $\rho^{{\rm tot}^{\scriptstyle \star}}_{e,\mathcal{g}'',\mathcal{h}} \geq \rho^{{\rm tot}^{\scriptstyle \star}}_{e,\mathcal{g}',\mathcal{h}}$;
        \item if $\mathcal{h}'' \geq \mathcal{h}'$, then $\rho^{{\rm tot}^{\scriptstyle \star}}_{e,\mathcal{g},\mathcal{h}''} \leq \rho^{{\rm tot}^{\scriptstyle \star}}_{e,\mathcal{g},\mathcal{h}'}$.
    \end{itemize}
    \begin{proof} 
        See Appendix~\ref{proof:OSP_increasing_decreasing}.
    \end{proof}
\end{thm}

This is an expected result, i.e., when the legitimate channel improves, then it is reasonable to use more energy in order to get a higher rate. Conversely, when the eavesdropper's channel improves, it is better not to use a lot of energy because only low rates can be obtained. In this case, it is better to conserve energy and wait for a better slot. The previous theorem is useful to prune the action space in the numerical computation: if we found the optimal transmission power for a given channel state, we could exploit it as lower [upper] bound for better [worse] channel states.

We expect that a result similar to Theorem~\ref{thm:OSP_increasing_decreasing} holds for a generic $N > 1$. A formal proof would require to explicitly compute $D(\rho^{\rm tot};\boldsymbol{\mathcal{g}}',\boldsymbol{\mathcal{h}}';\boldsymbol{\mathcal{g}}'',\boldsymbol{\mathcal{h}}'')$ and show that it is non-negative or non-positive (see Appendix~\ref{proof:OSP_increasing_decreasing}). However, this would require the computation of an analytical expression for $\eta$ in Equation~\eqref{eq:tau_r}. Even though this is in principle possible for any fixed $N$, the corresponding expression is very complicated and, in practice, the resulting $D(\rho^{\rm tot};\boldsymbol{\mathcal{g}}',\boldsymbol{\mathcal{h}}';\boldsymbol{\mathcal{g}}'',\boldsymbol{\mathcal{h}}'')$ is too long to be analytically tractable.

\section{Optimal Secrecy Policy with Partial CSI} \label{sec:statistical_CSI}

In the previous sections we assumed that the realizations of $\boldsymbol{G}$ and $\boldsymbol{H}$, namely $\boldsymbol{\mathcal{g}}$ and $\boldsymbol{\mathcal{h}}$, are known at the transmitter. This may not be true in practice. In particular, it is likely that, since the eavesdropper does not cooperate with the transmitter, its channel gain is unknown. In this section we gradually remove these assumptions and discuss how the achievable secrecy rate changes as a result.

We assume that $\boldsymbol{G} = [G_1,\ldots,G_N]$ and $\boldsymbol{H} = [H_1,\ldots,H_N]$ have independent components and are independent of each other. In this section  we assume that all links are affected by i.i.d. Nakagami fading. This means that the amplitude of a received signal has a Nakagami pdf with parameters $m$ and $\kappa$, i.e.,~
\begin{align}
    &f(x; m, \kappa) = 2 \left(\frac{m}{\kappa}\right)^m \frac{1}{\Gamma(m)} x^{2m-1} e^{-\frac{m}{\kappa}x^2},\quad x \geq 0,\\
    &\Gamma(m) \triangleq \int_0^\infty e^{-t}t^{m-1}\ \mbox{d}t.
\end{align}

Therefore, $G_r$ and $H_r$ exhibit a Gamma distribution. The pdf of $G_r$ (with mean $\bar{g}_r$) is~
\begin{align}\label{eq:f_G_i_x_m}
    f_{G_r}(\mathcal{g}; m) = \left(\frac{m}{\bar{g}_r}\right)^m\frac{1}{\Gamma(m)} \mathcal{g}^{m-1} e^{-\frac{m}{\bar{g}_r}\mathcal{g}}, \quad {\mathcal{g} \in \mathbb{R}_+,\atop m \geq 1}
\end{align}

\noindent and similarly for $H_r$ (for presentation simplicity, we assume that the legitimate receiver and the eavesdropper have the same index $m$, but the analysis can be extended to a more general case). Note that $m = 1$ corresponds to Rayleigh fading and $f_{G_r}(\mathcal{g};1) =  \frac{1}{\bar{g}_r}e^{-\mathcal{g}/\bar{g}_r}$ is an exponential distribution. As $m$ increases, the strength of the line of sight component increases. For ease of notation, in the remainder of the paper we drop the dependence on $m$ and implicitly assume $f_{G_r}(\mathcal{g}) = f_{G_r}(\mathcal{g}; m)$.

\subsection{Unknown Eavesdropper's Channel}\label{subsec:par_CSI}

In this section, we assume that both the legitimate and the eavesdropper's channels are affected by fading but CSI is available only for $\boldsymbol{G}$. In this case, due to this lack of information, it may happen that EHD transmits even when the eavesdropper's channel gain is higher than the legitimate one.

Similarly to Expression~\eqref{eq:mu_det} in the previous section, a policy $\mu$ can be defined as~
\begin{align}
    \mu = \{\boldsymbol{\rho}_{e,\boldsymbol{\mathcal{g}}} \triangleq [\rho_{1;e,\boldsymbol{\mathcal{g}}},\ldots,\rho_{N;e,\boldsymbol{\mathcal{g}}}] \in \mathcal{P}_{\scriptscriptstyle \leq}(e),\ \forall e \in \mathcal{E},\ \forall \boldsymbol{\mathcal{g}} \in \mathcal{G}\},
\end{align}

\noindent and similarly for $\mu^{\rm tot}$. $\boldsymbol{\rho}_{e,\boldsymbol{\mathcal{g}}}$ represents the transmission power used in state $(e,\boldsymbol{\mathcal{g}})$ (since $\boldsymbol{\mathcal{h}}$ is unknown, it cannot be included in the state of the system). We remark that $\mu$ performs a \emph{power control} mechanism, i.e., a policy specifies only the transmission power $\boldsymbol{\rho}_{e,\boldsymbol{\mathcal{g}}}$. However, in addition to power control, in every slot also the code rate can be changed (see Section~\ref{subsec:coding_strategy}). In particular, variable rate coding provides higher secrecy rates than constant rate coding, but is more difficult to implement. In the following we analyze both these approaches.\footnote{Differently from the complete CSI case of Section~\ref{sec:OSP_complete_CSI}, $\rho_r$ cannot be set to $0$ if $\mathcal{g}_r \leq \mathcal{h}_r$ (see Equation~\eqref{eq:tau_r}), thus using constant rate or variable rate coding leads to different results.}

\subsubsection{Constant Rate Coding}
The simplest assumption is that the coding scheme has constant rate and its choice only depends on the overall channel statistics. Using constant rate coding, the eavesdropper is able to gather more information than the legitimate receiver when its channel is better.
Because of this, for some $r$, we may have (see Equation~\eqref{eq:R_var_con})~
\begin{align}
  R_{\mathcal{g}_r,\mathcal{h}_r}(\rho_{r;e,\boldsymbol{\mathcal{g}}}) < 0.
\end{align}

The secrecy rate expression becomes~
\begin{align}\label{eq:C_mu_constant}
    \begin{split}
  C_{\mu} = & \sum_{e = 0}^{e_{\rm max}} \pi_{\mu^{\rm tot}}(e) \int_{\mathbb{R}_+^N}\int_{\mathbb{R}_+^N} \sum_{r = 1}^N \log_2\left(\frac{1+\mathcal{g}_r \rho_{r;e,\boldsymbol{\mathcal{g}}}}{1+\mathcal{h}_r \rho_{r;e,\boldsymbol{\mathcal{g}}}}\right) \\
  &  \times \prod_{r=1}^N \Big( f_{G_r}(\mathcal{g}_r)f_{H_r}(\mathcal{h}_r) \Big) \ \mbox{d}\boldsymbol{\mathcal{g}} \ \mbox{d}\boldsymbol{\mathcal{h}},
  \end{split}
\end{align}

Note that in~\eqref{eq:C_mu_constant} we integrate both positive and negative terms. The negative terms are due to the fact that the eavesdropper's channel may be better than the legitimate one ($\mathcal{h}_r > \mathcal{g}_r$).

We now want to extract some properties of the optimal secrecy policy in this context. We start by performing the following computations, which will be used to extend the first point of Theorem~\ref{thm:OSP_increasing_decreasing}.

The channel memoryless property can be used to simplify~\eqref{eq:C_mu_constant} and recast the problem using an MDP. By integrating over $\boldsymbol{\mathcal{h}}$, we obtain~
\begin{align}
    \begin{split}
      &C_{\mu} = \sum_{e = 0}^{e_{\rm max}} \pi_{\mu^{\rm tot}}(e) \int_{\mathbb{R}_+^N} \sum_{r = 1}^N T_r^{\rm con}(\mathcal{g}_r,\rho_{r;e,\boldsymbol{\mathcal{g}}}) \prod_{r=1}^N f_{G_r}(\mathcal{g}_r)  \ \mbox{d}\boldsymbol{\mathcal{g}}.
    \end{split} \label{eq:C_mu_con} \\
      &T_r^{\rm con}(\mathcal{g},\rho) \triangleq \int_{\mathbb{R}_+} \log_2\left(\frac{1+\mathcal{g} \rho}{1+\mathcal{h} \rho}\right) f_{H_r}(\mathcal{h}) \ \mbox{d}\mathcal{h}. \label{eq:T_m_con}
\end{align}

The function $T_r^{\rm con}(\mathcal{g},\rho)$ is presented in Equation~\eqref{eq:T_m_gamma_bar_h_omega}, where $\text{Ei}(z)=-\int_{-z}^\infty \frac{e^{-t}}{t} \ \mbox{d}t$ is the exponential integral function and $s_i$, $t_i$ are constants.\footnote{Closed form expressions for $s_i$ and $t_i$ can be derived but are quite complicated. Moreover, we will see that they do not contribute to our next results.}
\begin{align}\label{eq:T_m_gamma_bar_h_omega}
    \begin{split}
        T_r^{\rm con}(\mathcal{g},\rho) = & \ \log_2(1+\mathcal{g}\rho) + \frac{1}{\log 2}\sum_{i=2}^m \bigg(s_i \left(\rho \bar{h}_r\right)^{i-m} \\
        &  + e^{\frac{m}{\rho  \bar{h}_r}} \text{Ei}\left(-\frac{m}{\rho  \bar{h}_r}\right) \sum_{i=1}^m t_i \left(\rho \bar{h}_r\right)^{i-m}\bigg).
    \end{split}
\end{align}

A secure transmission can be performed only if $C_{\mu} > 0$. The maximum of \eqref{eq:C_mu_con} can be found with an MDP approach, where the MC state is given by the pair $(e,\boldsymbol{\mathcal{g}})$.

A property, that directly follows from the definitions of $T_r^{\rm con}(\mathcal{g},\rho)$, is the following.
\begin{propos}\label{propos:omega_infty}
    If for $\rho > 0$ we obtain $T_r^{\rm con}(\mathcal{g},\rho) < 0$, then allocating a power $\rho$ over sub-carrier $r$ is strictly sub-optimal.
\end{propos}

This result is intuitive. Indeed, if $T_r^{\rm con}(\mathcal{g},\rho) < 0$ and $\rho > 0$, then in~\eqref{eq:C_mu_con} we are adding negative terms. This is clearly sub-optimal because it lowers the secrecy rate and wastes energy at the same time.

Even if $T_r^{\rm con}(\mathcal{g},\rho)$ has a complicated expression, as we will see, we are interested in its double derivative with respect to $\mathcal{g}$ and $\rho$:~
\begin{align}\label{eq:der_con}
    \frac{\partial^2 }{\partial \rho \partial \mathcal{g}}T_r^{\rm con}(\mathcal{g},\rho) = \frac{1}{\log 2}\frac{1}{(1+\mathcal{g} \rho)^2}.
\end{align}

We now show that even with partial CSI the optimal secrecy policy increases with the legitimate channel gain. As for Theorem~\ref{thm:OSP_increasing_decreasing}, the following result can be used to prune the action space.\footnote{We provide a formal proof only for the case $N = 1$ because, even if theoretically possible, the proof for a generic $N > 1$ is not analytically tractable (see the related discussion just after Theorem~\ref{thm:OSP_increasing_decreasing}).}

\begin{thm}\label{thm:OSP_increasing_decreasing_partial_CSI}
    Consider $N = 1$. With partial CSI, the transmission power of OSP is non-decreasing with $\mathcal{g}$ (we omit the ``$1$'' subscripts). Formally, if $\mathcal{g}'' \geq \mathcal{g}'$, then $\rho^{{\rm tot}^{\scriptstyle \star}}_{e,\mathcal{g}''} \geq \rho^{{\rm tot}^{\scriptstyle \star}}_{e,\mathcal{g}'}$.
    \begin{proof}
        The proof follows the same steps presented in Appendices~\ref{proof:OSP_deterministic},~\ref{proof:D},~\ref{proof:OSP_increasing_decreasing}. To prove the theorem the key point is that~
        \begin{align}
            \frac{\partial^2}{\partial \rho \partial \mathcal{g}} T_r^{\rm con}(\mathcal{g},\rho) \geq 0. \label{eq:partial_2_T_m}
        \end{align}
        
        Note that, considering the derivative with respect to $\rho$, it follows from~\eqref{eq:partial_2_T_m} that $\frac{\partial}{\partial \mathcal{g}} T_r^{\rm con}(\mathcal{g},\rho_B) - \frac{\partial}{\partial \mathcal{g}} T_r^{\rm con}(\mathcal{g},\rho_A) \geq 0$, for $\rho_A \leq \rho_B$. We can rewrite the inequality as $\frac{\partial}{\partial \mathcal{g}} \big(T_r^{\rm con}(\mathcal{g},\rho_B)-T_r^{\rm con}(\mathcal{g},\rho_A)\big) \geq 0$ and obtain~
        \begin{align}
            \begin{split}
                &T_r^{\rm con}(\mathcal{g}+\Delta,\rho_A) - T_r^{\rm con}(\mathcal{g},\rho_A) \\
                &\leq T_r^{\rm con}(\mathcal{g}+\Delta,\rho_B) - T_r^{\rm con}(\mathcal{g},\rho_B),
            \end{split}
        \end{align}
        
        \noindent $\forall \Delta \geq 0$ and $\rho_A \leq \rho_B$. This condition can be replaced with Equation~\eqref{eq:ineq_3} in Appendix~\ref{proof:D} to prove the theorem.
    \end{proof}
\end{thm}

\subsubsection{Variable Rate Coding}

Better performance can be obtained with variable rate coding (see Equations~\eqref{eq:C_variable} and~\eqref{eq:C_constant}). In this case, in every slot, the code rate is matched to the legitimate channel rate. Thus, even if $\mathcal{g}_r \leq \mathcal{h}_r$ (eavesdropper's channel is better), the eavesdropper can gather at most $R_{\mathcal{g}_r}$ bits (legitimate transmission rate) and not $R_{\mathcal{h}_r}$ (eavesdropper's transmission rate).
The secrecy rate expression is~
\begin{align}\label{eq:C_mu_variable}
  C_{\mu} =& \sum_{e = 0}^{e_{\rm max}} \pi_{\mu^{\rm tot}}(e) \int_{\mathbb{R}_+^N}\int_{\mathbb{R}_+^N} \sum_{r = 1}^N \left[\log_2\left(\frac{1+\mathcal{g}_r \rho_{r;e,\boldsymbol{\mathcal{g}}}}{1+\mathcal{h}_r \rho_{r;e,\boldsymbol{\mathcal{g}}}}\right)\right]^+ \nonumber \\
  &  \times \prod_{r=1}^N \Big( f_{G_r}(\mathcal{g}_r)f_{H_r}(\mathcal{h}_r) \Big) \ \mbox{d}\boldsymbol{\mathcal{g}} \ \mbox{d}\boldsymbol{\mathcal{h}},
\end{align}

As before, we introduce a function $T_r^{\rm var}(\mathcal{g},\rho_{r;e,\boldsymbol{\mathcal{g}}})$ such that~
\begin{align}
  C_{\mu} =& \sum_{e = 0}^{e_{\rm max}} \pi_{\mu^{\rm tot}}(e) \int_{\mathbb{R}_+^N} \sum_{r = 1}^N T_r^{\rm var}(\mathcal{g}_r,\rho_{r;e,\boldsymbol{\mathcal{g}}}) \prod_{r=1}^N f_{G_r}(\mathcal{g}_r)  \ \mbox{d}\boldsymbol{\mathcal{g}}. \label{eq:C_mu_var}
\end{align}
\vspace{-\belowdisplayskip}
\vspace{-\abovedisplayskip}
\begin{align}
    T_r^{\rm var}(\mathcal{g},\rho) &\triangleq \int_{\mathbb{R}_+} \left[\log_2\left(\frac{1+\mathcal{g} \rho}{1+\mathcal{h} \rho}\right)\right]^+ f_{H_r}(\mathcal{h}) \ \mbox{d}\mathcal{h} \\
  &= \int_{0}^{\mathcal{g}} \log_2\left(\frac{1+\mathcal{g} \rho}{1+\mathcal{h} \rho}\right) f_{H_r}(\mathcal{h}) \ \mbox{d}\mathcal{h}. \label{eq:T_m_var}
\end{align}

In Equation~\eqref{eq:T_m_var} we integrate from zero to $\mathcal{g}$, thus we remove the $[\cdot]^+$ notation (see the structure of Equation~\eqref{eq:R_var_con} with variable rate coding).

Note that $T_r^{\rm var}(\mathcal{g},\rho) \geq T_r^{\rm con}(\mathcal{g},\rho)$, which justifies the fact that the achievable secrecy rate with variable rate coding is higher than with constant rate coding.

The analogous of Theorem~\ref{thm:OSP_increasing_decreasing_partial_CSI} holds in this case, as can be proved by exploiting the structure of the double derivative of $T_r^{\rm var}(\mathcal{g},\rho)$:~
\begin{align}\label{eq:der_var}
    \frac{\partial^2}{\partial \rho \partial \mathcal{g}} T_r^{\rm var}(\mathcal{g},\rho) = \frac{1}{\log 2}\frac{\Gamma (m)-\Gamma \left(m,\frac{\textstyle  m \mathcal{g} }{\textstyle \bar{h}_r}\right)}{(1+\mathcal{g}  \rho)^2 \Gamma (m)},
\end{align}

\noindent where $\Gamma(m,z) \triangleq \int_z^\infty e^{-t} t^{m-1} \ \mbox{d}t$ is the incomplete gamma function.

\subsection{No Channel State Information}\label{subsec:no_CSI}

Lower secrecy rates are obtained when also the legitimate receiver's channel is unknown. In particular, the transmission power cannot be adapted to the current channel state. It is easy to show that $C_{\mu}$ can be greater than zero only if $\bar{g}_r > \bar{h}_r$ for some $r$. However, the mean values of the channel gains are not controlled by the transmitter (they are physical quantities), thus if the legitimate channel is (statistically) worse, no secrecy can be achieved.

\begin{figure}[t]
  \centering
  \includegraphics[trim = 4mm 0mm 4mm 5mm,  clip, width=1\columnwidth, height=.75\columnwidth]{omega_e.eps}
  \vspace{-0.49cm}
  \caption{Transmission power $\rho^{{\rm tot}^{\scriptstyle \star}}_{e,\mathcal{g},\mathcal{h}}$ as a function of the battery level $e$ for several values of $\mathcal{h}$ and $\mathcal{g} \in [0.41,0.51)$.}
  \label{fig:omega_e}
\end{figure}

\section{Numerical Evaluation}\label{sec:numerical_evaluation}
In this section we discuss how the secrecy rate changes as a function of the different system parameters.

We compare the following scenarios: OSP with full CSI (OSP-FULL), OSP with only legitimate channel knowledge and constant rate coding (OSP-PAR-CON) or variable rate coding (OSP-PAR-VAR) and OSP with only statistical channel knowledge (OSP-STAT).

If not otherwise stated, the simulation parameters are: $e_{\rm max} = 30$, truncated geometric energy arrivals with $b_{\rm max} = 6$ and $\bar{b} = 1$, $n = 15$ quantization intervals (see Section~\ref{sec:finite_model}), $N = 1$ (single sub-carrier), $\bar{g} = \bar{h} = 1$ (symmetric scenario), $\mathcal{G} = \mathcal{H} = \mathbb{R}_+$ with $m = 1$ (Rayleigh fading). After showing results for this choice of parameters, we study the sensitivity of the system performance by changing one or more parameters while keeping the others fixed.

\subsubsection{\underline{Fixed Parameters}}

\figurename~\ref{fig:omega_e} shows the optimal transmission power $\rho^{{\rm tot}^{\scriptstyle \star}}_{e,\mathcal{g},\mathcal{h}}$ as a function of the battery level $e$ when $\mathcal{g} \in [0.41,0.51)$ and $\mathcal{h} \in \mathbb{R}_+$. We recall that, when $\mathcal{G} = \mathcal{H} = \mathbb{R}_+$, we use the technique explained in Section~\ref{sec:finite_model}, i.e., we have a finite number of points where the transmission power is computed ($n = 15$). When $\mathcal{h} \geq 0.51$, the transmission power is identically zero because the eavesdropper is always advantaged. Also when $\mathcal{h} \in [0.41,0.51)$ the transmission power is zero. This is not obvious a priori and strongly depends upon the considered interval of $\mathcal{g}$. It can be seen that Theorem~\ref{thm:OSP_increasing_decreasing} holds, i.e., $\rho^{{\rm tot}^{\scriptstyle \star}}_{e,\mathcal{g},\mathcal{h}}$ does not increase with $\mathcal{h}$. Finally, we note that the behavior of the transmission power is not obvious a priori, e.g., it is significantly different from a simple greedy policy ($\rho^{{\rm tot}^{\scriptstyle \star}}_{e,\mathcal{g},\mathcal{h}} = e$) even when $\mathcal{h}$ is low.

\begin{figure}[t]
  \centering
  \includegraphics[trim = 4mm 0mm 4mm 5mm,  clip, width=1\columnwidth, height=.75\columnwidth]{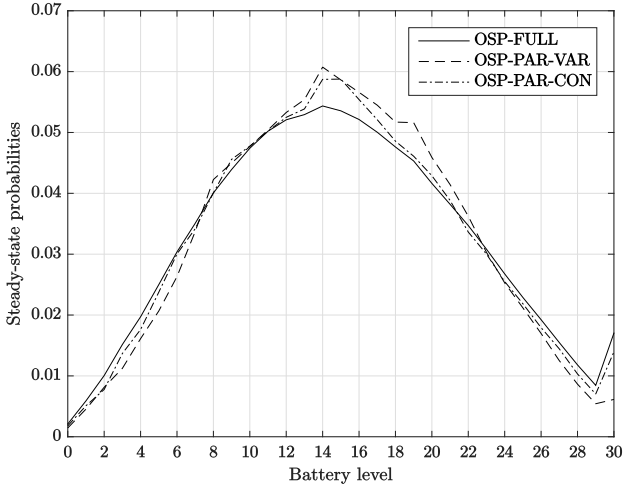}
  \caption{Steady-state probabilities $\pi_{\mu^{\rm tot}}(e)$ as a function of the battery level $e$.}
  \label{fig:pi_e}
\end{figure}

\figurename~\ref{fig:pi_e}, instead, shows the steady-state probabilities as a function of the energy level $e$, for fixed $e_{\rm max}$ and in the different scenarios. In all cases, the curves are similar. This is because the device tends to operate in an efficient region, i.e., approximately at $e_{\rm max}/2$. This is in order to avoid energy outage and overflow, that degrade the performance of the system. When $e$ approaches $e_{\rm max}$, the steady-state tails increase because of the overflow (when the battery is almost full, all harvesting events leading to overflow contribute to increasing the steady-state probability of state $e_{\rm max}$, which is then higher than those of the immediately lower states).

\subsubsection{\underline{Battery Size}}

\begin{figure}[t]
  \centering
  \includegraphics[trim = 4mm 0mm 4mm 5mm,  clip, width=1\columnwidth, height=.75\columnwidth]{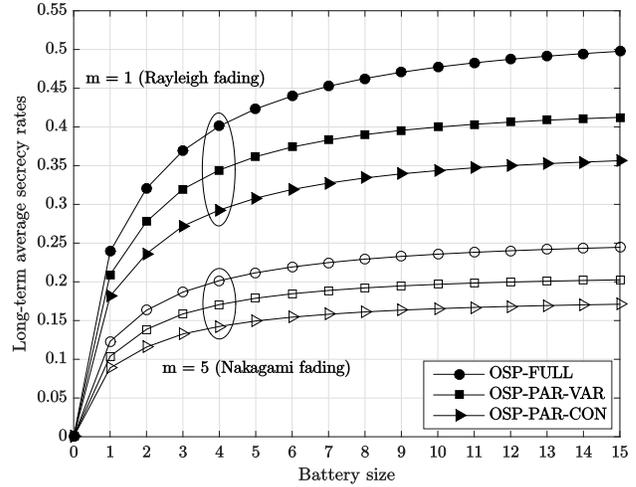}
  \caption{Secrecy rate $C_\mu$ as a function of the battery size $e_{\rm max}$ in the case of symmetric channel conditions.}
  \label{fig:plot_e_max_symmetric}
\end{figure}

In \figurename~\ref{fig:plot_e_max_symmetric} we show the rate achieved by the various policies as a function of the battery size $e_{\rm max}$. We use Rayleigh ($m = 1$) and a general Nakagami fading with a strong Line of Sight (LoS) component ($m = 5$). The curves of OSP-STAT are identically zero because $\bar{g} = \bar{h}$. As expected, OSP-FULL has the highest secrecy rate for every value. It can be seen that the curves saturate after a certain value. This is due to the combination of two effects: 1) the harvesting rate of the EHD is limited (it can be shown that the performance of an EH system is bounded) and 2) the achievable secrecy rate always saturates in the high power regime (because of the structure of Equation~\eqref{eq:R_var_con}). Note that the curves saturate already for small $e_{\rm max}$, therefore, in practice, it may be sufficient to use small batteries to obtain high secrecy rates.

\begin{figure}[t]
  \centering
  \includegraphics[trim = 4mm 0mm 4mm 5mm,  clip, width=1\columnwidth, height=.75\columnwidth]{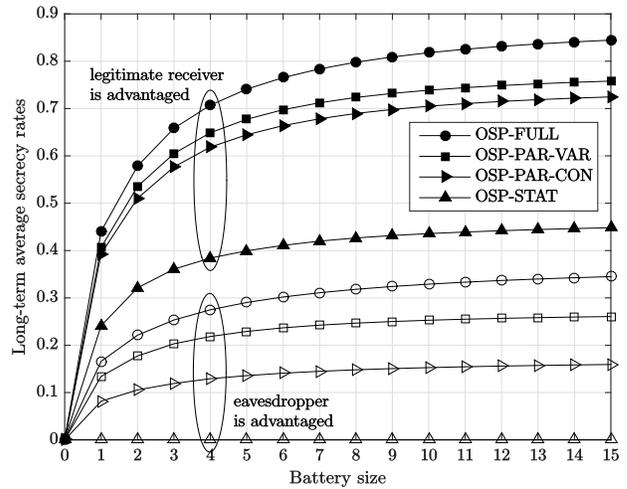}
  \caption{Secrecy rate $C_\mu$ as a function of the battery size $e_{\rm max}$ in the case of asymmetric channel conditions and Rayleigh fading.}
  \label{fig:plot_e_max_asymmetric}
\end{figure}
In~\cite[Section~IV-B]{Gopala2008} the authors showed that, when the transmission is subject to an average power constraint, the performance of the optimal transmission scheme with variable rate coding and partial CSI knowledge approaches the performance of the full CSI case when the transmission power is sufficiently high. In our previous example, OSP-PAR-VAR does not achieve OSP-FULL when $e_{\rm max}$ increases because an energy harvesting system imposes an average power constraint $\bar{b}$.\footnote{This can be easily derived starting from the causality constraint \begin{align}
    \sum_{k = 0}^K \sum_{r = 1}^N \Sigma_r^{(k)} \leq E^{(0)} + \sum_{k = 0}^{K-1} B^{(k)}, \qquad \forall K = 0,1,\ldots
\end{align}

\noindent where, according to Equation~\eqref{eq:Ep1_E}, $\Sigma_r^{(k)}$ is the transmission power over sub-carrier $r$ in time slot $k$, $B^{(k)}$ is the amount of energy harvested in slot $k$ and $E^{(0)}$ is the amount of energy initially available in the battery. In the long run, the right-hand side becomes the power constraint of our system.} It can be verified that, when $\bar{b}$ increases, if the battery size is sufficiently large, the gap between OSP-PAR-VAR and OSP-FULL is smaller.

Note that the achievable secrecy rates strongly depend upon the fading statistics. With $m = 5$, we have strong LoS components, i.e., the channel pdfs tend to be narrow around their means ($\bar{g} = \bar{h}$). It follows that the legitimate and eavesdropper's channel gains are close to each other most of the time. This corresponds to low values of $R_{\mathcal{g}_r,\rho_r}(\rho_r)$, thus a low secrecy rate. With Rayleigh fading, instead, exploiting channel diversity allows to obtain higher rewards. This is also the reason why, with Rayleigh fading, full channel state information  (OSP-FULL) provides a great improvement with respect to the partial knowledge cases.

\figurename~\ref{fig:plot_e_max_asymmetric} is similar to the previous one but with asymmetric channel gains. When the eavesdropper is advantaged ($\bar{g} = 1$, $\bar{h} = 2$), even if low performance can be achieved, secret transmission is still possible. When OSP-PAR-CON is used, it is likely that EHD transmits even when the eavesdropper's channel is better and in this case, from Equation~\eqref{eq:C_mu_con}, the secrecy rate is lower. This effect is emphasized if the eavesdropper's channel is advantaged, because it is more likely that the legitimate channel is the worse of the two.

On the other hand, if the legitimate channel is better ($\bar{g} = 2$, $\bar{h} = 1$), the secrecy rate can reach high values. In this case, OSP-STAT is also considered and, as expected, is the worst among the optimal policies.

\begin{figure}[t]
  \centering
  \includegraphics[trim = 4mm 0mm 4mm 5mm,  clip, width=1\columnwidth, height=.75\columnwidth]{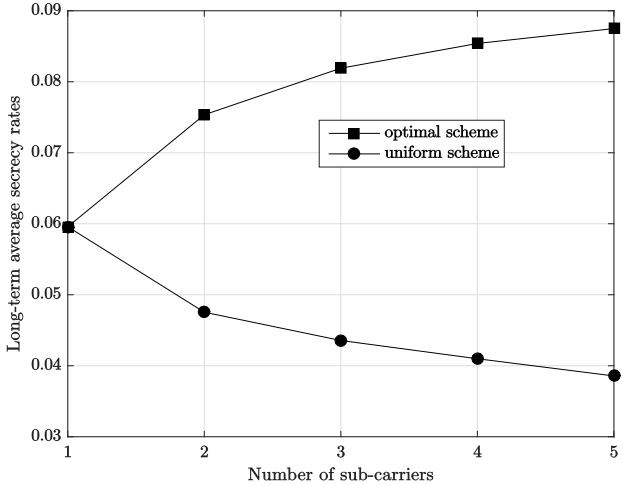}
  \caption{Secrecy rate $C_\mu$ as a function of the number of sub-carriers $N$.\newline}
  \label{fig:plot_N}
\end{figure}
\subsubsection{\underline{Number of sub-carriers}}

When $N = 1$, finding the optimal policies for high values of $n$ (fine quantization of the channel gains) is feasible. We recall that the number of states of the MC is directly proportional to the number of possible combinations of channel gains. Thus, with $N = 1$, the possible combinations are $n\times n$ (legitimate channel $\times$ eavesdropper's channel). With a generic $N$, the combinations become $n^N \times n^N$. Thus, the number of states grows exponentially with the number of sub-carriers, making the optimization process for high $N$ infeasible in practice (curse-of-dimensionality). Even when the problem symmetry can be exploited (when $G_r$ and $H_r$ are i.i.d.), the computational effort still remains heavy. In pratice, this approach can be applied to multi-carrier scenarios if the number of carriers, $N$, and the number of quantization levels for the channel, $n$, are not too large. Note however that our solution suffers from a dimensionality problem because it is the \emph{optimal} solution. Part of our future work agenda includes the design of sub-optimal schemes and the study of trade-offs between computational times and performance.

\begin{figure}[t]
  \centering
  \includegraphics[trim = 4mm 0mm 4mm 5mm,  clip, width=1\columnwidth, height=.75\columnwidth]{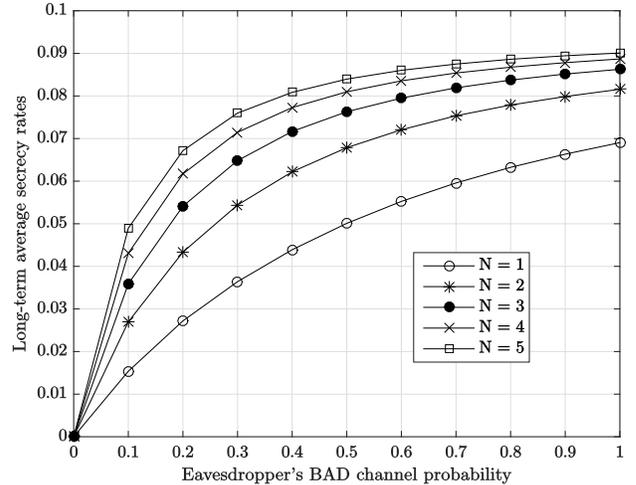}
  \caption{Secrecy rate $C_\mu$ of OSP-FULL as a function of the eavesdropper's BAD channel probability in a binary channel system.}
  \label{fig:plot_N_BAD}
\end{figure}

In the following, as an example, we consider a discrete GOOD-BAD channel and discuss the importance of the power splitting scheme. We define $\mathcal{G} = \mathcal{H} = \{{\rm B},{\rm G}\} = \{1/30,3/30\} = \{-15 \mbox{ dB},-10 \mbox{ dB}\}$ with probabilities $0.7$ and $0.3$, respectively. We also set $e_{\rm max} = 10$ because, generally, the saturation region is almost reached for this battery size (see~\figurename s~\ref{fig:plot_e_max_symmetric} and~\ref{fig:plot_e_max_asymmetric}). In \figurename~\ref{fig:plot_N}, we plot OSP-FULL as a function of the number of sub-carriers $N$ when the optimal (Equations~\eqref{eq:tau_r}-\eqref{eq:alpha_beta}) or a uniform power splitting is used. In the optimal case, as $N$ increases, the reward also increases. This is expected because, when one user experiences a bad channel condition, then the power can be directed to other good sub-carriers. Instead, with uniform power splitting, the secrecy rate decreases with $N$. In practice, this happens because, instead of sending all the transmission power in the ``good'' sub-carriers, a fraction of this is wasted in the ``bad'' sub-carriers. For example, with $N = 2$, it may happen that over sub-carrier $1$ the pair legitimate-eavesdropper's channel gain is $(G,B)$ whereas, for sub-carrier $2$, the pair is $(B,B)$, i.e., sub-carrier $1$ is a ``good'' sub-carrier while sub-carrier $2$ is not. In this case, if a positive transmission power were used, the corresponding reward would be greater than zero but the power sent over sub-carrier $2$ would be wasted (only when the two pairs are $(G,B)$ and $(G,B)$, is no power wasted during the transmission). This explains why the performance degrades as the number of sub-carriers increases. Moreover, the effect is emphasized with larger $N$ because there are more cases where the transmission power cannot be fully exploited.

When the legitimate and the eavesdropper's channel gains are known in every slot, using a smart power splitting scheme is convenient because it can significantly improve the network performance. 
If this is not possible (e.g., because this information is not available or not reliable), a sub-optimal strategy needs to be adopted, e.g., uniform power splitting, which is simpler to implement but yields lower performance in general. The study of the information/performance tradeoff for power splitting strategies is left for future work.

Finally, \figurename~\ref{fig:plot_N_BAD} shows how the optimal secrecy rate changes as a function of $\mathbb{P}(h_1 = {\rm B}) = \mathbb{P}(h_2 = {\rm B}) \in [0,1]$ for different numbers of sub-carriers. It can be noticed that the case with five sub-carriers and $\mathbb{P}(h_1 = {\rm B}) = 0.2$ achieves the same performance as the system with only one sub-carrier but $\mathbb{P}(h_1 = {\rm B}) = 1$. In practice, the diversity offered by a greater number of sub-carriers can be efficiently exploited to obtain higher secrecy rates. 
An interesting point is that, as $N$ increases, the improvement obtained from $N$ to $N+1$ decreases. This is due to the concavity properties of Equation~\eqref{eq:c_rho_c_r}. Therefore, it may not be necessary to use a large number of sub-carriers to obtain high secrecy rates.

\section{Conclusions} \label{sec:conclusions}
In this work we analyzed an Energy Harvesting Device that has a finite energy storage and transmits secret data to a receiver over $N$ parallel channels exploiting physical layer characteristics. 
We found the best power allocation technique, namely the Optimal Secrecy Policy (OSP), in several contexts depending on the degree of channel knowledge  the device has. We proved several properties of OSP and in particular that it is deterministic and monotonic. We also described a technique to compute OSP by decomposing the problem in two steps and using a dynamic programming approach. When only partial channel state information is available, we described how the maximum secrecy rate varies with constant and variable rate coding, explaining and numerically evaluating the advantages of variable rate coding. We numerically showed that, because of the limited harvesting rate that is inherently provided by the renewable energy source, OSP-PAR-VAR does not achieve the same performance of OSP-FULL as the battery size increases, and noted that it is not necessary to use very large batteries to achieve close to optimal performance. We also set up the problem when more than one sub-carrier is considered, and discussed the scalability problems related to such scenario. Also, we found that using the optimal power splitting scheme provides a significant advantage with respect to the simpler uniform splitting approach.

Future work may include the study of sub-optimal strategies for the case with $N$ sub-carriers in order to avoid the curse-of-dimensionality problem. Also, other optimization techniques can be investigated, e.g., offline approach, Lyapunov optimization or reinforcement learning approach. Finally, it would be interesting to set up a simulation experiment with real data measurements (e.g., for the harvesting process) in order to validate our results in a realistic scenario.

\appendices

\section{Proof of Theorem~\ref{thm:OSP_deterministic}} \label{proof:OSP_deterministic}

We want to show that OSP is a deterministic policy, i.e., given the state of the system, $\mu(\boldsymbol{\rho} ; e,\boldsymbol{\mathcal{g}},\boldsymbol{\mathcal{h}}) = \delta_{\boldsymbol{\rho},\boldsymbol{\rho}_{e,\boldsymbol{\mathcal{g}},\boldsymbol{\mathcal{h}}}^\star}$, where $\delta_{\cdot,\cdot}$ is the Kronecker delta function.\footnote{A proof of this result in the discounted horizon case can be found in~\cite[Theorems~6.2.9 and~6.2.10]{Puterman1995}. In our discussion we follow a different approach which will also be useful to prove Proposition~\ref{propos:D}.}

Note that the study can be split into two parts according to Equation~\eqref{eq:mu_decomposition}. Thus, we only need to prove that both $\gamma_\mu(\rho^{\rm tot} ; e,\boldsymbol{\mathcal{g}},\boldsymbol{\mathcal{h}})$ (transmit power policy) and $\phi_\mu(\boldsymbol{\rho};\rho^{\rm tot},e,\boldsymbol{\mathcal{g}},\boldsymbol{\mathcal{h}})$ are deterministic. In the following we prove the first part. The latter is derived in~\cite{Gopala2008}.

\subsection{Deterministic Transmit Power Policy}

As a preliminary result, we need the following proposition (in this subsection, the expectation is always taken with respect to~$\boldsymbol{G}$ and~$\boldsymbol{H}$).

\begin{propos}
    $\mathbb{P}(E^{(k)} = e | E^{(0)})$ depends upon the policy only through $\mathbb{E}[\gamma_\mu(\rho^{\rm tot} ; e,\boldsymbol{G},\boldsymbol{H})]$, $\forall \rho^{\rm tot} \in \{0,\ldots,e\}$, $\forall e \in \mathcal{E}$.
    \begin{proof}
        The proof is by induction on $k$.
        At $k = 0$, $\mathbb{P}(E^{(0)} = e | E^{(0)} = e_0)$ is equal to $1$ if $e = e_0$ and to $0$ otherwise. In this case there is no dependence upon the policy.

        Assume that the thesis is true for $k$ (inductive hypothesis). Using the chain rule, the probability that $E^{(k+1)} = e'$ given the initial state is~
        \begin{align}
            \mathbb{P}(E^{(k+1)} = e' | E^{(0)}) =& \sum_{e = 0}^{e_{\rm max}} \mathbb{P}(E^{(k+1)} = e' | E^{(k)} = e) \\ 
            &\times \mathbb{P}(E^{(k)} = e | E^{(0)}).\nonumber
        \end{align}
        
        Thus, to prove the thesis, we focus on $\mathbb{P}(E^{(k+1)} = e' | E^{(k)} = e)$, whereas for $\mathbb{P}(E^{(k)} = e | E^{(0)})$ we use the inductive hypothesis.
        Assume $e' < e_{\rm max}$~
        \begin{align}
            &\mathbb{P}(E^{(k+1)} = e' | E^{(k)} = e) \\ 
            &=\sum_{b = \max\{0,e'-e\}}^{\min\{e',b_{\rm max}\}} p_B(b) \mathbb{E}[\gamma_\mu(e-e'+b ; e,\boldsymbol{G},\boldsymbol{H})], \nonumber
        \end{align}
        \noindent whereas, if $e' = e_{\rm max}$~
        \begin{align}
            &\mathbb{P}(E^{(k+1)} = e_{\rm max} | E^{(k)} = e) \\ 
            &=\sum_{b = \max\{0,e_{\rm max}-e\}}^{b_{\rm max}} p_B(b) \sum_{d = 0}^{e-e_{\rm max}+b}\mathbb{E}[\gamma_\mu(d ; e,\boldsymbol{G},\boldsymbol{H})].\nonumber
        \end{align}
        
        Note that we used the \emph{transmit power policy} $\gamma_\mu(\cdot)$ and not the \emph{power allocation policy} $\mu(\cdot)$. Indeed, the battery evolution does not depend upon the particular power splitting scheme but only on the total energy consumed.
        Thus, $\mathbb{P}(E^{(k+1)} = e' | E^{(0)})$ depends upon the policy only through the expectations $\mathbb{E} [\gamma_\mu(\rho^{\rm tot} ; E^{(k)},\boldsymbol{G},\boldsymbol{H})]$.
    \end{proof}
\end{propos}

Define now the long-term probabilities of being in the energy level $e$ given the initial level $E^{(0)}$ as $\pi(e | E^{(0)}) = \liminf_{K \rightarrow \infty} \frac{1}{K+1} \sum_{k = 0}^K \mathbb{P}(E^{(k)} = e | E^{(0)})$. Thanks to the above proposition, we know that $\pi(e | E^{(0)})$ depends upon the policy only through $\mathbb{E} [\gamma_\mu(\rho^{\rm tot} ; e,\boldsymbol{G},\boldsymbol{H})]$, $\forall \rho^{\rm tot} \in \{0,\ldots,e\}$, $\forall e \in \mathcal{E}$.

Fix a value $\alpha(\rho^{\rm tot};e)$ for every pair $\rho^{\rm tot}$ and $e$, and consider the set of policies $\Xi$ that induce $\mathbb{E} [\gamma_\mu(\rho^{\rm tot} ; e,\boldsymbol{G},\boldsymbol{H})] = \alpha(\rho^{\rm tot};e)$ for every pair. For every policy in $\Xi$, the long-term probabilities are the same. The long-term average secrecy rate given an initial state $E^{(0)}$ can be expressed as in Equation~\eqref{eq:C_mu_E^{(0)}}~
\begin{align}
    C_\mu(E^{(0)}) =& \sum_{e \in \mathcal{E}} \pi(e | E^{(0)}) \\
    &\times \mathbb{E}\bigg[\sum_{\boldsymbol{\rho} \in \mathcal{P}_{\scriptscriptstyle \leq}(e)}\mu(\boldsymbol{\rho};e,\boldsymbol{G},\boldsymbol{H}) c(\boldsymbol{\rho},\boldsymbol{G},\boldsymbol{H}) \bigg].\nonumber
\end{align}

For every policy in $\Xi$, the terms $\pi(e | E^{(0)})$ of the previous expression are the same. Therefore, in order to maximize $C_\mu(E^{(0)})$, we focus on the terms $\mathbb{E}[\cdot]$ for each value of $e$. In particular, the problem can be decomposed in $e_{\rm max}+1$ simpler optimization problems (according to~\eqref{eq:mu}, define $\mu(e) \triangleq \{\mu(\cdot;e,\boldsymbol{\mathcal{g}},\boldsymbol{\mathcal{h}}),\ \forall \boldsymbol{\mathcal{g}} \in \mathcal{G}, \ \boldsymbol{\mathcal{h}} \in \mathcal{H}\}$)~
\begin{subequations}
\begin{align}
    \begin{split}
        \max_{\mu(e)} \ \mathbb{E} \bigg[\sum_{\boldsymbol{\rho} \in \mathcal{P}_{\scriptscriptstyle \leq}(e)}\mu(\boldsymbol{\rho};e,\boldsymbol{G},\boldsymbol{H}) c(\boldsymbol{\rho},\boldsymbol{G},\boldsymbol{H}) \bigg],
    \end{split}
\end{align}
\vspace{-\belowdisplayskip}
\vspace{-\abovedisplayskip}
\begin{alignat}{2}
\shortintertext{s.t.:}
	&\mbox{Constraints in }\eqref{eq:mu_constraints}; \\
    &\mathbb{E} [\gamma_\mu(\rho^{\rm tot} ; e,\boldsymbol{G},\boldsymbol{H})] = \alpha(\rho^{\rm tot};e), \ && \forall \rho^{\rm tot} \in \{0,\ldots,e\}.
\end{alignat}
\label{eq:proof_det_opt_problem}
\end{subequations}

We rewrite the first expression as follows~
\begin{align}\label{eq:proof_det_max_simpl}
    \max_{\mu(e)} \ \mathbb{E} \bigg[&\sum_{\rho^{\rm tot} \in \{0,\ldots,e\}}\gamma_\mu(\rho^{\rm tot};e,\boldsymbol{G},\boldsymbol{H}) \\
    &\times \!\!\!\!\sum_{\boldsymbol{\rho} \in \mathcal{P}_{\scriptscriptstyle =}(\rho^{\rm tot})}\phi_\mu(\boldsymbol{\rho};\rho^{\rm tot},e,\boldsymbol{\mathcal{g}},\boldsymbol{\mathcal{h}}) c(\boldsymbol{\rho},\boldsymbol{G},\boldsymbol{H}) \bigg]. \nonumber
\end{align}

\noindent where $\mathcal{P}_{\scriptscriptstyle =}(\rho^{\rm tot}) \triangleq \{\boldsymbol{\rho}\ : \ \boldsymbol{\rho} \succeq 0,\ \rho^{\rm tot} = \sum_{r=1}^N \rho_r \}$. As derived in~\cite[Eq.~7]{Gopala2008} with a Lagrangian approach, $\phi_\mu(\boldsymbol{\rho};\rho^{\rm tot},e,\boldsymbol{\mathcal{g}},\boldsymbol{\mathcal{h}}) = \delta_{\boldsymbol{\rho},\boldsymbol{\tau}_{\rho^{\rm tot},\boldsymbol{\mathcal{g}},\boldsymbol{\mathcal{h}}}^\star}$ ($\phi_\mu(\cdot)$ is deterministic and there is no dependence upon $e$ when $\rho^{\rm tot}$ is fixed). $\boldsymbol{\tau}_{\rho^{\rm tot},\boldsymbol{\mathcal{g}},\boldsymbol{\mathcal{h}}}^\star$ is the optimal transmit power splitting given the total transmission power $\rho^{\rm tot}$ and the channel gains (we use $\boldsymbol{\tau}$ instead of $\boldsymbol{\rho}$ for notation clarity). Therefore, we can rewrite~\eqref{eq:proof_det_max_simpl} as~
\begin{align}
    \max_{\gamma_\mu(e)} \ \mathbb{E} \bigg[&\sum_{\rho^{\rm tot} \in \{0,\ldots,e\}}\gamma_\mu(\rho^{\rm tot};e,\boldsymbol{G},\boldsymbol{H}) c(\boldsymbol{\tau}_{\rho^{\rm tot},\boldsymbol{G},\boldsymbol{H}}^\star,\boldsymbol{G},\boldsymbol{H}) \bigg].
\end{align}

For every fixed $e$, we want to define $\gamma_\mu(e) \triangleq \{\gamma_\mu(\cdot;e,\boldsymbol{\mathcal{g}},\boldsymbol{\mathcal{h}}),\ \forall \boldsymbol{\mathcal{g}} \in \mathcal{G}, \ \boldsymbol{\mathcal{h}} \in \mathcal{H}\}$. Note that the problem is concave, thus a Lagrangian approach can be used. The Lagrangian function is
\begin{align}
        \mathcal{L}(e) =& \ \mathbb{E} \bigg[\sum_{\rho^{\rm tot} \in \{0,\ldots,e\}} \gamma_\mu(\rho^{\rm tot};e,\boldsymbol{G},\boldsymbol{H}) \label{eq:L_e} \\
        & \times \Big(c(\boldsymbol{\tau}_{\rho^{\rm tot},\boldsymbol{G},\boldsymbol{H}}^\star,\boldsymbol{G},\boldsymbol{H}) - \lambda(\rho^{\rm tot};e) \Big)\bigg],\nonumber
\end{align}

\noindent where $\lambda(\rho^{\rm tot};e)$ is the Lagrange multiplier associated with constraint $\mathbb{E} [\gamma_\mu(\rho^{\rm tot} ; e,\boldsymbol{G},\boldsymbol{H})] = \alpha(\rho^{\rm tot};e)$.

We now show that an optimal policy is $\gamma_\mu(\rho^{\rm tot} ; e,\boldsymbol{\mathcal{g}},\boldsymbol{\mathcal{h}}) = 1$ if $\rho^{\rm tot} = {\rho^{\rm tot}_{e,\boldsymbol{\mathcal{g}},\boldsymbol{\mathcal{h}}}}^\star$ and zero otherwise, with~
\begin{align}\label{eq:omega_star_det}
    {\rho^{\rm tot}_{e,\boldsymbol{\mathcal{g}},\boldsymbol{\mathcal{h}}}}^\star = \argmax{\rho^{\rm tot} \in \{0,\ldots,e\}} \Big\{ c(\boldsymbol{\tau}_{\rho^{\rm tot},\boldsymbol{\mathcal{g}},\boldsymbol{\mathcal{h}}}^\star,\boldsymbol{\mathcal{g}},\boldsymbol{\mathcal{h}}) - \lambda(\rho^{\rm tot};e) \Big\}.
\end{align}

In order to maximize~\eqref{eq:L_e}, we can focus on each argument of the expectation~
\begin{align}
        \max_{\substack{\gamma_\mu(\rho^{\rm tot};e,\boldsymbol{\mathcal{g}},\boldsymbol{\mathcal{h}}),\\ \forall \rho^{\rm tot} \in \{0,\ldots,e\}}} \sum_{\rho^{\rm tot} \in \{0,\ldots,e\}} & \gamma_\mu(\rho^{\rm tot};e,\boldsymbol{\mathcal{g}},\boldsymbol{\mathcal{h}}) \label{eq:weighted_sum} \\
        &\times \underbrace{\Big(c(\boldsymbol{\tau}_{\rho^{\rm tot},\boldsymbol{\mathcal{g}},\boldsymbol{\mathcal{h}}}^\star,\boldsymbol{\mathcal{g}},\boldsymbol{\mathcal{h}}) - \lambda(\rho^{\rm tot};e) \Big)}_{u(\rho^{\rm tot},e,\boldsymbol{\mathcal{g}},\boldsymbol{\mathcal{h}})}. \nonumber
\end{align}

We recall that $\sum_{\rho^{\rm tot} \in \{0,\ldots,e\}} \gamma_\mu(\rho^{\rm tot} ; e,\boldsymbol{\mathcal{g}},\boldsymbol{\mathcal{h}}) = 1$. \eqref{eq:weighted_sum} is a weighted sum that is maximized when $\gamma_\mu(\rho^{\rm tot} ; e,\boldsymbol{\mathcal{g}},\boldsymbol{\mathcal{h}}) = 1$ if $\rho^{\rm tot} = {\rho^{\rm tot}_{e,\boldsymbol{\mathcal{g}},\boldsymbol{\mathcal{h}}}}^\star$ and zero otherwise. Indeed, suppose by contradiction that there exist $\rho^{\rm tot}_1$ and $\rho^{\rm tot}_2$ (the argument can be generalized to more than two) such that $\gamma_\mu(\rho^{\rm tot}_1 ; e,\boldsymbol{\mathcal{g}},\boldsymbol{\mathcal{h}})>0$, $\gamma_\mu(\rho^{\rm tot}_2 ; e,\boldsymbol{\mathcal{g}},\boldsymbol{\mathcal{h}})>0$ and $\gamma_\mu(\rho^{\rm tot}_1 ; e,\boldsymbol{\mathcal{g}},\boldsymbol{\mathcal{h}})+\gamma_\mu(\rho^{\rm tot}_2 ; e,\boldsymbol{\mathcal{g}},\boldsymbol{\mathcal{h}}) = 1$. The max argument in~\eqref{eq:weighted_sum} would be $\gamma_\mu(\rho^{\rm tot}_1 ; e,\boldsymbol{\mathcal{g}},\boldsymbol{\mathcal{h}})u(\rho^{\rm tot}_1,e,\boldsymbol{\mathcal{g}},\boldsymbol{\mathcal{h}})+(1-\gamma_\mu(\rho^{\rm tot}_1 ; e,\boldsymbol{\mathcal{g}},\boldsymbol{\mathcal{h}}))u(\rho^{\rm tot}_2,e,\boldsymbol{\mathcal{g}},\boldsymbol{\mathcal{h}})$, which is smaller than or equal to $u({\rho^{\rm tot}_{e,\boldsymbol{\mathcal{g}},\boldsymbol{\mathcal{h}}}}^\star,e,\boldsymbol{\mathcal{g}},\boldsymbol{\mathcal{h}})$.

\section{Proof of Proposition~\ref{propos:irreducible}} \label{proof:irreducible}
The MC has three dimensions: the battery, the legitimate channel and the eavesdropper's channel. Since the fading is not controlled by the EHD, the MC is always free to move along the last two dimensions (we assume that the channel evolution is i.i.d. over time). Thus, the only potential problem is related to the battery dimension, i.e., if the policy is not unichain, the device energy level may be stuck in different subsets of $\mathcal{E}$.

        Also, we recall that we consider only discrete channel conditions with non-zero probability (Remark~\ref{remark:non_null}). We now discuss Point~1). We want to show that the recurrent class is composed by the states with high energy levels, i.e., for every $e < e_{\rm max}$, there exists a positive probability of increasing the energy level. This is true by hypothesis because the maximum transmit power in state $e$ is lower than the maximum number of energy arrivals $b_{\rm max}$ ($\rho^{\rm tot}_{e,\boldsymbol{\mathcal{g}}',\boldsymbol{\mathcal{h}}'} < b_{\rm max}$). Therefore, since it is possible to reach the energy level $e_{\rm max}$ (fully charged battery) within a certain number of steps from every state, the policy is unichain.
To prove Point~2), a symmetric reasoning can be followed.
        
        If both conditions hold, it is possible to reach every $e \in \mathcal{E}$ from any element of $\mathcal{E}$, thus the policy induces an irreducible MC. Since the number of states is finite, the MC is positive recurrent.

\section{Deriving a Unichain Policy} \label{app:derive_unichain}

As in Appendix~\ref{proof:irreducible}, it is always possible to move along the channel dimensions. Therefore, we focus on the battery dimension, which represents the only limitation for obtaining a unichain policy.

Consider a policy $\mu_A$ that has two recurrent classes, namely $\Pi_A'$ and $\Pi_A''$ (this approach can be generalized to more than two classes) and assume, without loss of generality, that if $E^{(0)} \in \Pi_A''$ the greatest long-term reward is reached. We now propose a technique to derive a new policy that, regardless of the initial state, achieves the same maximum reward of $\mu_A$.

Consider a second policy, namely $\mu_B$, obtained from $\mu_A$ as follows. For every $e_A = 0,\ldots,\max\{\Pi_A''\}$, set $\boldsymbol{\rho}_{e_B,\boldsymbol{\mathcal{g}},\boldsymbol{\mathcal{h}}}^{\mu_B} = \boldsymbol{\rho}_{e_A,\boldsymbol{\mathcal{g}},\boldsymbol{\mathcal{h}}}^{\mu_A}$, with $e_B = e_A + e_{\rm max} - \max_e\{\Pi_A''\}$, i.e., we shift the recurrent class $\Pi_A''$ toward higher energy levels (we name $\Pi_B''$ the new recurrent class). For $e_B \in \{0,\ldots,e_{\rm max}-\max\{\Pi_A''\}-1\}$, set $\boldsymbol{\rho}_{e_B,\boldsymbol{\mathcal{g}},\boldsymbol{\mathcal{h}}}^{\mu_B} = 0$.
In this way, the device cannot be stuck in energy levels lower than $e_{\rm max}-|\Pi_B''|+1$ (the harvested energy increases the battery level) and, after a certain number of transitions, it reaches the recurrent class $\Pi_B''$.
Finally, since the power splitting vectors in the recurrent classes $\Pi_A''$ and $\Pi_B''$ coincide, $\mu_B$ achieves the same maximum reward of $\mu_A$, regardless of the initial $E^{(0)}$.

This proves that it is always possible to obtain a unichain policy with the same maximum long-term secrecy rate as the initial one and shows how to derive it.

\section{Proof of Theorem~\ref{thm:max_C_Omega}} \label{proof:max_C_Omega}

Problem~\eqref{eq:max_problem} can be rewritten using~\eqref{eq:mu_tot_det} in the following form:~
\begin{align}
	&\max_\mu C_\mu = \max_{\mu^{\rm tot}} \max_{\mu \in \mathcal{X}(\mu^{\rm tot})} C_\mu \\
	&\mathcal{X}(\mu^{\rm tot}) \triangleq \{\mu \ : \ \mu^{\rm tot} \mbox{ and } \mu \mbox{ are consistent}\},
\end{align}

\noindent i.e., we fix the transmission powers (outer $\max$) and focus on all the policies which are consistent with such choice (inner $\max$). This is equivalent to searching through all the possible feasible policies (as in~\eqref{eq:max_problem}).
  
Consider the expression of $C_\mu$ in Equation~\eqref{eq:C_fading_simplified} and note that $\pi_{\mu^{\rm tot}}(e)$ does not depend upon the particular power splitting scheme, but only upon $\mu^{\rm tot}$. Thus, the inner $\max$ can be moved inside the integral~
\begin{align}
    \max_{\mu^{\rm tot}}\bigg(& \sum_{e = 0}^{e_{\rm max}} \pi_{\mu^{\rm tot}}(e) \\
    & \times \int_{\mathcal{G}\times \mathcal{H}} \max_{\mu \in \mathcal{X}(\mu^{\rm tot})}\Big(c(\boldsymbol{\rho}_{e,\boldsymbol{\mathcal{g}},\boldsymbol{\mathcal{h}}},\boldsymbol{\mathcal{g}},\boldsymbol{\mathcal{h}}) \Big)  \ \mbox{d}F (\boldsymbol{\mathcal{g}},\boldsymbol{\mathcal{h}}) \bigg).\nonumber
\end{align}
    
Note that inside the integral $e$, $\boldsymbol{\mathcal{g}}$ and $\boldsymbol{\mathcal{h}}$ are fixed. Therefore, the only degree of freedom in the inner $\max$ operation is given by the power splitting choice $\boldsymbol{\rho}_{e,\boldsymbol{\mathcal{g}},\boldsymbol{\mathcal{h}}}$.

Since $\mu^{\rm tot}$ and $\mu$ are consistent, in the inner $\max$ we have $\boldsymbol{\rho}_{e,\boldsymbol{\mathcal{g}},\boldsymbol{\mathcal{h}}} \in \mathcal{P}_{\scriptscriptstyle =}({\rho^{\rm tot}_{e,\boldsymbol{\mathcal{g}},\boldsymbol{\mathcal{h}}}})$ (specified in~\eqref{eq:sigma_star}). Therefore,~
\begin{align}
    \max_{\mu \in \mathcal{X}(\mu^{\rm tot})}\Big(c(\boldsymbol{\rho}_{e,\boldsymbol{\mathcal{g}},\boldsymbol{\mathcal{h}}},\boldsymbol{\mathcal{g}},\boldsymbol{\mathcal{h}}) \Big) \equiv \mbox{Problem \eqref{eq:sigma_star} with } x = {\rho^{\rm tot}_{e,\boldsymbol{\mathcal{g}},\boldsymbol{\mathcal{h}}}}
\end{align}

Thus, Points 1) and 2) of the theorem solve the internal and external $\max$ operations, respectively.
    
\section{Proof of Proposition~\ref{propos:D}}\label{proof:D}

The proof exploits the results of Appendix~\ref{proof:OSP_deterministic}, and in particular Equation~\eqref{eq:omega_star_det}. Also, we focus on the energy levels in the unique recurrent class (for the transient states the proposition is trivial to prove since $\rho^{{\rm tot}^{\scriptstyle \star}}_{e,\boldsymbol{\mathcal{g}}',\boldsymbol{\mathcal{h}}'}$ is always zero).

Assume that $\rho^{{\rm tot}'} \triangleq \rho^{{\rm tot}^{\scriptstyle \star}}_{e,\boldsymbol{\mathcal{g}}',\boldsymbol{\mathcal{h}}'}$ is the optimal transmission power given the state of the system $(e,\boldsymbol{\mathcal{g}}',\boldsymbol{\mathcal{h}}')$, i.e., $\rho^{{\rm tot}^{\scriptstyle \star}}_{e,\boldsymbol{\mathcal{g}}',\boldsymbol{\mathcal{h}}'} = \arg \max_{\rho^{\rm tot} \in \{0,\ldots,e\}} \{ c(\boldsymbol{\tau}_{\rho^{\rm tot},\boldsymbol{\mathcal{g}}',\boldsymbol{\mathcal{h}}'}^\star,\boldsymbol{\mathcal{g}}',\boldsymbol{\mathcal{h}}') - \lambda(\rho^{\rm tot};e) \}$ (we remark that $\boldsymbol{\tau}_{\rho^{\rm tot},\boldsymbol{\mathcal{g}}',\boldsymbol{\mathcal{h}}'}^\star$ is the \emph{optimal} power splitting vector given $\rho^{\rm tot}$ and the channel gains). Similarly, $\rho^{\rm tot ''} \triangleq \rho^{{\rm tot}^{\scriptstyle \star}}_{e,\boldsymbol{\mathcal{g}}'',\boldsymbol{\mathcal{h}}''}$ is the optimal power for state $(e,\boldsymbol{\mathcal{g}}'',\boldsymbol{\mathcal{h}}'')$. 

We first show by contradiction that if $D(\rho^{\rm tot};\boldsymbol{\mathcal{g}}',\boldsymbol{\mathcal{h}}';\boldsymbol{\mathcal{g}}'',\boldsymbol{\mathcal{h}}'') \geq 0, \forall \rho^{\rm tot}$, then $\rho^{\rm tot ''} \geq \rho^{\rm tot '}$. Assume $\rho^{\rm tot '} > \rho^{\rm tot ''}$.
We now derive some properties of $\rho^{\rm tot '}$ and $\rho^{\rm tot ''}$ and combine these with the hypothesis to obtain the contradiction.
From the definitions of $\rho^{\rm tot '}$ and $\rho^{\rm tot ''}$, we have~
\begin{align}
    \begin{split}
        &c(\boldsymbol{\tau}_{\rho^{\rm tot '},\boldsymbol{\mathcal{g}}',\boldsymbol{\mathcal{h}}'}^\star,\boldsymbol{\mathcal{g}}',\boldsymbol{\mathcal{h}}') - \lambda(\rho^{\rm tot '};e) \\
        &\qquad \qquad \geq c(\boldsymbol{\tau}_{\rho^{\rm tot ''},\boldsymbol{\mathcal{g}}',\boldsymbol{\mathcal{h}}'}^\star, \boldsymbol{\mathcal{g}}',\boldsymbol{\mathcal{h}}') - \lambda(\rho^{\rm tot ''};e),
    \end{split}\label{eq:ineq_1}\\
    \begin{split}
        &c(\boldsymbol{\tau}_{\rho^{\rm tot ''},\boldsymbol{\mathcal{g}}'',\boldsymbol{\mathcal{h}}''}^\star,\boldsymbol{\mathcal{g}}'',\boldsymbol{\mathcal{h}}'') - \lambda(\rho^{\rm tot ''};e) \\ 
        &\qquad \qquad \geq c(\boldsymbol{\tau}_{\rho^{\rm tot '},\boldsymbol{\mathcal{g}}'',\boldsymbol{\mathcal{h}}''}^\star, \boldsymbol{\mathcal{g}}'',\boldsymbol{\mathcal{h}}'') - \lambda(\rho^{\rm tot '};e).
    \end{split}\label{eq:ineq_2}
\end{align}

By hypothesis, we have, for every $\rho^{\rm tot}$,~
\begin{align}\label{eq:D_proof}
    \frac{\partial}{\partial \rho^{\rm tot}}\Big(c(\boldsymbol{\tau}_{\rho^{\rm tot},\boldsymbol{\mathcal{g}}'',\boldsymbol{\mathcal{h}}''}^\star,\boldsymbol{\mathcal{g}}'',\boldsymbol{\mathcal{h}}'') - c(\boldsymbol{\tau}_{\rho^{\rm tot},\boldsymbol{\mathcal{g}}',\boldsymbol{\mathcal{h}}'}^\star,\boldsymbol{\mathcal{g}}',\boldsymbol{\mathcal{h}}')\Big) \geq 0.
\end{align}

Assume that the inequality is strict. This implies, for every $\rho_A < \rho_B$~
\begin{align}
    \begin{split}
        &c(\boldsymbol{\tau}_{\rho_A,\boldsymbol{\mathcal{g}}'',\boldsymbol{\mathcal{h}}''}^\star,\boldsymbol{\mathcal{g}}'',\boldsymbol{\mathcal{h}}'') - c(\boldsymbol{\tau}_{\rho_A,\boldsymbol{\mathcal{g}}',\boldsymbol{\mathcal{h}}'}^\star,\boldsymbol{\mathcal{g}}',\boldsymbol{\mathcal{h}}') \\
        & < c(\boldsymbol{\tau}_{\rho_B,\boldsymbol{\mathcal{g}}'',\boldsymbol{\mathcal{h}}''}^\star,\boldsymbol{\mathcal{g}}'',\boldsymbol{\mathcal{h}}'') - c(\boldsymbol{\tau}_{\rho_B,\boldsymbol{\mathcal{g}}',\boldsymbol{\mathcal{h}}'}^\star,\boldsymbol{\mathcal{g}}',\boldsymbol{\mathcal{h}}').
    \end{split}
\end{align}

In particular, since $\rho^{\rm tot '} > \rho^{\rm tot ''}$, choose $\rho_A = \rho^{\rm tot ''}$ and $\rho_B = \rho^{\rm tot '}$ and obtain~
\begin{align}\label{eq:ineq_3}
    \begin{split}
        &c(\boldsymbol{\tau}_{\rho^{\rm tot ''},\boldsymbol{\mathcal{g}}'',\boldsymbol{\mathcal{h}}''}^\star,\boldsymbol{\mathcal{g}}'',\boldsymbol{\mathcal{h}}'') - c(\boldsymbol{\tau}_{\rho^{\rm tot ''},\boldsymbol{\mathcal{g}}',\boldsymbol{\mathcal{h}}'}^\star,\boldsymbol{\mathcal{g}}',\boldsymbol{\mathcal{h}}') \\
        & < c(\boldsymbol{\tau}_{\rho^{\rm tot '},\boldsymbol{\mathcal{g}}'',\boldsymbol{\mathcal{h}}''}^\star,\boldsymbol{\mathcal{g}}'',\boldsymbol{\mathcal{h}}'') - c(\boldsymbol{\tau}_{\rho^{\rm tot '},\boldsymbol{\mathcal{g}}',\boldsymbol{\mathcal{h}}'}^\star,\boldsymbol{\mathcal{g}}',\boldsymbol{\mathcal{h}}').
    \end{split}
\end{align}

Finally, by combining~\eqref{eq:ineq_2} with~\eqref{eq:ineq_3}, we obtain~
\begin{align}
        \begin{split}        
        &c(\boldsymbol{\tau}_{\rho^{\rm tot '},\boldsymbol{\mathcal{g}}'',\boldsymbol{\mathcal{h}}''}^\star, \boldsymbol{\mathcal{g}}'',\boldsymbol{\mathcal{h}}'') - \lambda(\rho^{\rm tot '};e) + \lambda(\rho^{\rm tot ''};e) \\
        &   \leq c(\boldsymbol{\tau}_{\rho^{\rm tot ''},\boldsymbol{\mathcal{g}}'',\boldsymbol{\mathcal{h}}''}^\star,\boldsymbol{\mathcal{g}}'',\boldsymbol{\mathcal{h}}'') \\
        &   < c(\boldsymbol{\tau}_{\rho^{\rm tot '},\boldsymbol{\mathcal{g}}'',\boldsymbol{\mathcal{h}}''}^\star,\boldsymbol{\mathcal{g}}'',\boldsymbol{\mathcal{h}}'') - c(\boldsymbol{\tau}_{\rho^{\rm tot '},\boldsymbol{\mathcal{g}}',\boldsymbol{\mathcal{h}}'}^\star,\boldsymbol{\mathcal{g}}',\boldsymbol{\mathcal{h}}') \\
        & \quad + c(\boldsymbol{\tau}_{\rho^{\rm tot ''},\boldsymbol{\mathcal{g}}',\boldsymbol{\mathcal{h}}'}^\star,\boldsymbol{\mathcal{g}}',\boldsymbol{\mathcal{h}}'),
        \end{split}
\end{align}

\noindent which is equivalent to~
\begin{align}
    \begin{split}
        & c(\boldsymbol{\tau}_{\rho^{\rm tot '},\boldsymbol{\mathcal{g}}',\boldsymbol{\mathcal{h}}'}^\star,\boldsymbol{\mathcal{g}}',\boldsymbol{\mathcal{h}}') - \lambda(\rho^{\rm tot '};e) \\
        &   < c(\boldsymbol{\tau}_{\rho^{\rm tot ''},\boldsymbol{\mathcal{g}}',\boldsymbol{\mathcal{h}}'}^\star,\boldsymbol{\mathcal{g}}',\boldsymbol{\mathcal{h}}') - \lambda(\rho^{\rm tot ''};e),
    \end{split} \label{eq:proof_combination1}
\end{align}

\noindent and violates Equation~\eqref{eq:ineq_1}, leading to a contradiction.

Assume now that~\eqref{eq:D_proof} holds with equality. Following the previous reasoning, we obtain~
\begin{align}
    \begin{split}
        &c(\boldsymbol{\tau}_{\rho^{\rm tot ''},\boldsymbol{\mathcal{g}}'',\boldsymbol{\mathcal{h}}''}^\star,\boldsymbol{\mathcal{g}}'',\boldsymbol{\mathcal{h}}'') - c(\boldsymbol{\tau}_{\rho^{\rm tot ''},\boldsymbol{\mathcal{g}}',\boldsymbol{\mathcal{h}}'}^\star,\boldsymbol{\mathcal{g}}',\boldsymbol{\mathcal{h}}') \\
        & \pushright{= c(\boldsymbol{\tau}_{\rho^{\rm tot '},\boldsymbol{\mathcal{g}}'',\boldsymbol{\mathcal{h}}''}^\star,\boldsymbol{\mathcal{g}}'',\boldsymbol{\mathcal{h}}'') - c(\boldsymbol{\tau}_{\rho^{\rm tot '},\boldsymbol{\mathcal{g}}',\boldsymbol{\mathcal{h}}'}^\star,\boldsymbol{\mathcal{g}}',\boldsymbol{\mathcal{h}}')}
    \end{split}
\end{align}

\noindent and, instead of~\eqref{eq:proof_combination1},~
\begin{align}
    \begin{split}
        & c(\boldsymbol{\tau}_{\rho^{\rm tot '},\boldsymbol{\mathcal{g}}',\boldsymbol{\mathcal{h}}'}^\star,\boldsymbol{\mathcal{g}}',\boldsymbol{\mathcal{h}}') - \lambda(\rho^{\rm tot '};e) \\
        &  \leq c(\boldsymbol{\tau}_{\rho^{\rm tot ''},\boldsymbol{\mathcal{g}}',\boldsymbol{\mathcal{h}}'}^\star,\boldsymbol{\mathcal{g}}',\boldsymbol{\mathcal{h}}') - \lambda(\rho^{\rm tot ''};e),
    \end{split}\label{eq:proof_combination2}
\end{align}

\eqref{eq:proof_combination2} must be satisfied with equality, otherwise it would violate~\eqref{eq:ineq_1}. This means that, for the same state $(e,\boldsymbol{\mathcal{g}}',\boldsymbol{\mathcal{h}}')$, there exist two distinct values of $\rho^{\rm tot}$ (i.e., $\rho^{\rm tot '}$ and $\rho^{\rm tot ''}$) that maximize~\eqref{eq:omega_star_det}. This is not possible because in the recurrent states the optimal solution is unique~\cite[Vol. II, Sec. 4]{Bertsekas2005}.

The first point of Proposition~\ref{propos:D} is thus proved. The proof of the second point is symmetric.

\section{Proof of Theorem~\ref{thm:OSP_increasing_decreasing}}\label{proof:OSP_increasing_decreasing}

We want to prove that, for OSP and $N = 1$, $\rho^{{\rm tot}^{\scriptstyle \star}}_{e,\mathcal{g},\mathcal{h}}$ does not decrease with $\mathcal{g}$ and does not increase with $\mathcal{h}$.

$D(\rho^{\rm tot};\mathcal{g}',\mathcal{h}';\mathcal{g}'',\mathcal{h}'')$ can be written as~
\begin{align}
    &D(\rho^{\rm tot};\mathcal{g}',\mathcal{h};\mathcal{g}'',\mathcal{h}) \\
    &= \frac{\partial}{\partial \rho^{\rm tot}}\left(\!\left[\log_2\!\left(\frac{1 + \mathcal{g}'' \rho^{\rm tot}}{1 + \mathcal{h} \rho^{\rm tot}} \right)\right]^+ \!\!\!\! - \left[\log_2\!\left(\frac{1 + \mathcal{g}' \rho^{\rm tot}}{1 + \mathcal{h} \rho^{\rm tot}} \right)\right]^+\right).\nonumber
\end{align}

Assume $\mathcal{g}'' \geq \mathcal{g}'$. If $\mathcal{g}''\leq \mathcal{h}$, then both terms are zero because $\mathcal{g}' \leq \mathcal{g}'' \leq \mathcal{h}$. If $\mathcal{g}' \leq \mathcal{h} < \mathcal{g}''$, then only the right term is zero. In this case, $D(\rho^{\rm tot};\mathcal{g}',\mathcal{h};\mathcal{g}'',\mathcal{h}) \propto \mathcal{g}''-\mathcal{h} > 0$. If $\mathcal{h} < \mathcal{g}' \leq \mathcal{g}''$, then $D(\rho^{\rm tot};\mathcal{g}',\mathcal{h};\mathcal{g}'',\mathcal{h}) \propto \mathcal{g}'' - \mathcal{g}' \geq 0$.

The proof of the second part is similar.

\bibliography{bibliography}{}

\bibliographystyle{IEEEtran}

\begin{IEEEbiography}[{\includegraphics[width=1in,height=1.25in,clip,keepaspectratio]{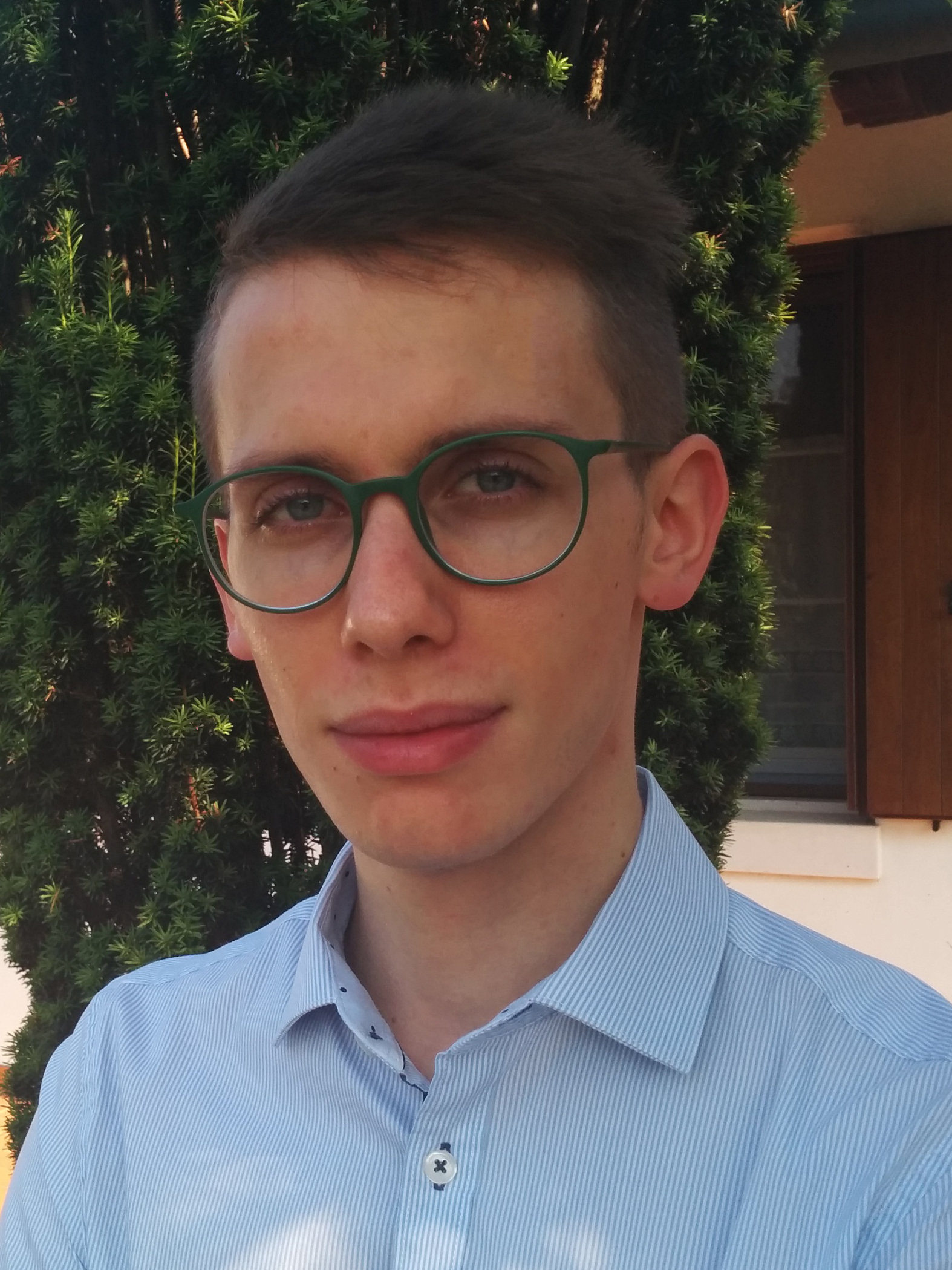}}]{Alessandro Biason}(S'15) received the B.Sc. degree (with honors) in Information Engineering and the M.S. degree (with honors and perfect GPA) in Telecommunication Engineering from the University of Padua, Italy, in 2012 and 2014, respectively. In 2015, he was on leave at the University of Southern California, Los Angeles, USA, as a visiting Ph.D. student. He is currently pursuing the Ph.D. degree with the SIGNET Research Group, University of Padua. His research interests lie in the areas of communication theory, wireless networks, energy harvesting systems, stochastic optimization and optimal control.
  
  He has served as a reviewer for the \textsc{IEEE Transactions on Communications}, \textsc{IEEE Transactions on Wireless Communications},  \textsc{IEEE Transactions on Mobile Computing} and \textsc{IEEE Journal on Selected Areas in Communications}.
\end{IEEEbiography}

\begin{IEEEbiography}[{\includegraphics[width=1in,height=1.25in,clip,keepaspectratio]{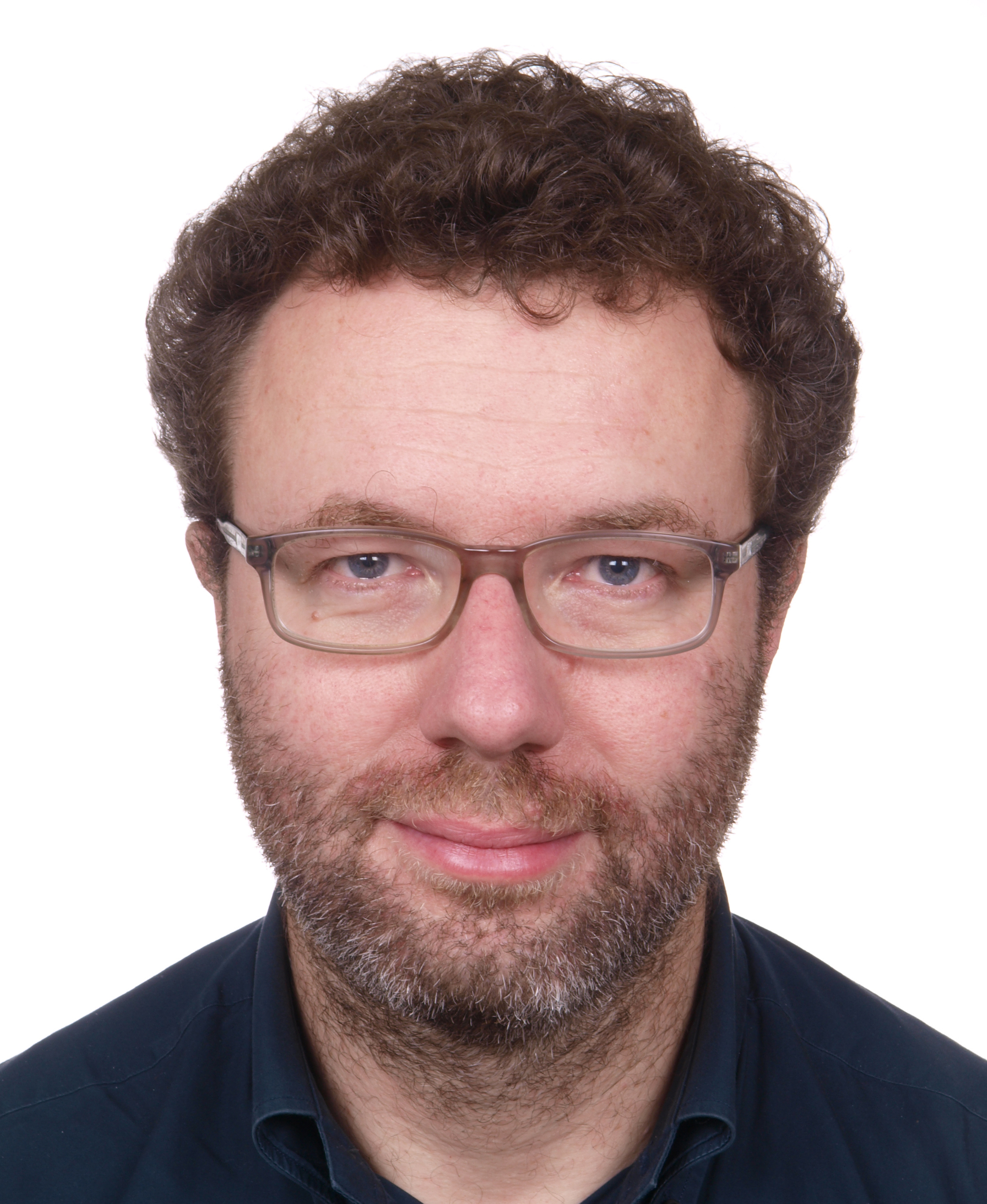}}]{Nicola Laurenti} received his Laurea Degree in Electrical Engineering in 1995 and his PhD in Electronic and Telecommunication Engineering in 1999 both from the University of Padua, Italy. Since 2001 he has been an Assistant Professor at the Department of Information Engineering of University of Padua. In 2008-09 he was a Visiting Scholar at the Coordinated Science Laboratory of the University of Illinois at Urbana-Champaign. In 1992-93 he was an exchange student at the University of California at Berkeley.
His research interests mainly focus on wireless network security at lower layers (physical, data link and network), GNSS security, information theoretic security and quantum key distribution.

\end{IEEEbiography}

\begin{IEEEbiography}[{\includegraphics[width=1in,height=1.25in,clip,keepaspectratio]{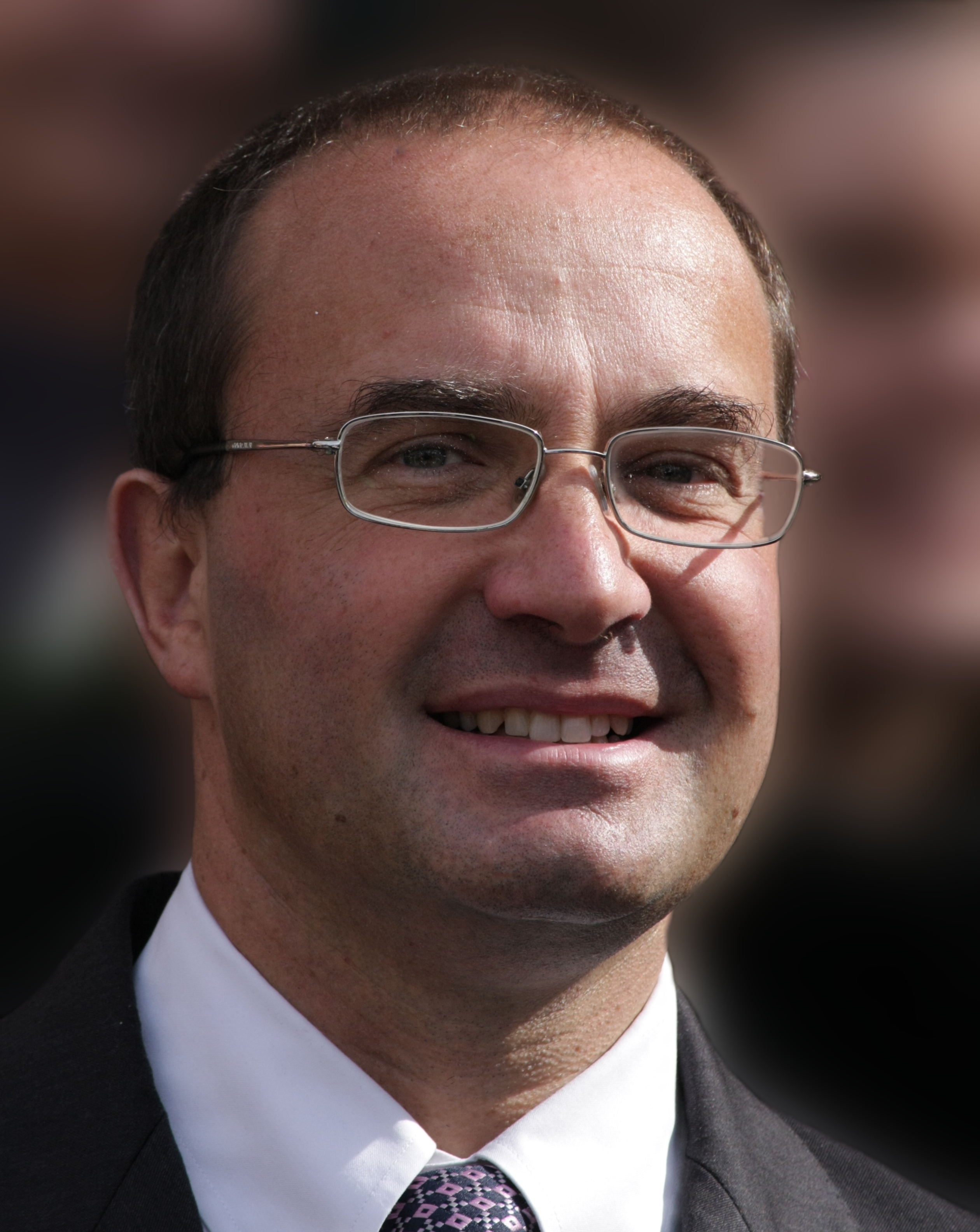}}]{Michele Zorzi}(S'89, M'95, SM'98, F'07) received his Laurea and PhD degrees in electrical engineering from the University of Padova in 1990 and 1994, respectively. During academic year 1992-1993 he was on leave at UCSD, working on multiple access in mobile radio networks. In 1993 he joined the faculty of the Dipartimento di Elettronica e Informazione, Politecnico di Milano, Italy. After spending three years with the Center for Wireless Communications at UCSD, in 1998 he joined the School of Engineering of the University of Ferrara, Italy, where he became a professor in 2000. Since November 2003 he has been on the faculty of the Information Engineering Department at the University of Padova. His present research interests include performance evaluation in mobile communications systems, random access in wireless networks, ad hoc and sensor networks, Internet-of-Things, energy constrained communications protocols, cognitive networks, and underwater communications and networking.

He was the Editor-In-Chief of IEEE Wireless Communications from 2003 to 2005 and the Editor-In-Chief of the IEEE Transactions on Communications from 2008 to 2011, and is currently the founding Editor-In-Chief of the \textsc{IEEE Transactions on Cognitive Communications and Networking}. He has also been an Editor for several journals and a member of the Organizing or the Technical Program Committee for many international conferences, as well as guest editor for special issues in IEEE Personal Communications, IEEE Wireless Communications, IEEE Network and the \textsc{IEEE Journal on Selected Areas in Communications}. He served as a Member-at-Large of the Board of Governors of the IEEE Communications Society from 2009 to 2011, and as its Director of Education and Training in 2014-15. He currently serves as a member of the 2016 IEEE PSPB/TAB Products and Services Committee.
\end{IEEEbiography}

\end{document}